\documentclass[aps,pra,reprint,superscriptaddress,footinbib,]{revtex4-2}

\usepackage[T1]{fontenc}				
\usepackage[utf8]{inputenc}			
\usepackage[english]{babel}			
\usepackage[protrusion=false]{microtype}	

\usepackage{xcolor} 						
\definecolor{red}{rgb}{1,0,0}				
\definecolor{blue}{rgb}{0,0,1}				
\definecolor{black}{rgb}{0,0,0}				

\usepackage{soul} 
\definecolor{hlyellow}{rgb}{0.95,0.95,0}
\definecolor{hlgreen}{rgb}{0,0.95,0}
\sethlcolor{hlyellow}
\soulregister \ref 7
\soulregister \cite 7
\soulregister \citet 7
\soulregister \footnote 8
\soulregister {\ } 7
\soulregister \figref 7
\soulregister \tabref 7
\soulregister \secref 7
\soulregister \appref 7
\soulregister \refref 7
\soulregister \eqref 7
\soulregister \supmat 7

\usepackage{amsmath} 
\usepackage{amssymb,amsfonts,mathrsfs} 
\usepackage{array} 
\usepackage{upgreek} 
\usepackage{braket}
\usepackage{bbold}

\usepackage{graphicx} 
\usepackage{makecell}
\usepackage{longtable}
\usepackage{booktabs}
\usepackage{placeins} 

\usepackage{comment}
\usepackage{hyperref} 
\definecolor{dullmagenta}{rgb}{0.4,0,0.4} 
\definecolor{darkblue}{rgb}{0,0,0.4}
\definecolor{medblue}{rgb}{0,0,0.6}
\definecolor{lightblue}{rgb}{0,0,0.8}
\hypersetup{
	colorlinks=true, 		
	breaklinks=true,		
	linkcolor=lightblue,
	citecolor=lightblue,
	urlcolor=lightblue,
	filecolor=dullmagenta
}
\usepackage[all]{hypcap} 

\newcommand{\figletter}[1]{\textbf{(\mbox{#1})}}
\newcommand{\figref}[1]{Fig.~\ref{#1}} 
\newcommand{\figrefs}[2]{Figs.~\ref{#1}-\ref{#2}} 
\newcommand{\tabref}[1]{Tab.~\ref{#1}} 
\newcommand{\secref}[1]{Sec.~\ref{#1}} 
\newcommand{\appref}[1]{Appendix \ref{#1}} 
\newcommand{\refref}[1]{Ref.~\cite{#1}} 
\newcommand{\eqnref}[1]{Eq.~\eqref{#1}} 

\newcommand{\GeSiGe}{Ge/SiGe} 

\newcommand{\muB}{\mu_\text{B}}
\newcommand{\gtens}{$g$-tensor} 
\newcommand{\gtenss}{$g$-tensors}
\newcommand{\gfact}{$g$-factor}
\newcommand{\gfacts}{$g$-factors}
\newcommand{\gtenspdf}{g-tensor}
\newcommand{\geff}{\ensuremath{g^*}}
\newcommand{\STm}{ST$_{-}$}

\newcommand{\Stwozero}{\ensuremath{\ket{\text{S}(2,0)}}}
\newcommand{\Tmoneone}{\ensuremath{\ket{\text{T}_{-}(1,1)}}}

\newcommand{\ma}[1]{\begin{align} #1 \end{align}}  

\newcommand{\vct}[1]{\boldsymbol{#1}}				
\newcommand{\uvct}[1]{\hat{\boldsymbol{#1}}}		
\newcommand{\mtrx}[1]{\boldsymbol{\mathbf{#1}}}		
\newcommand{\tens}{\mtrx}				
\newcommand{\of}[1]{\left( #1 \right)} 		
\newcommand{\D}{\mathrm{d}}					
\newcommand{\Exp}[1]{\exp \left[#1\right]}  		
\newcommand{\abs}[1]{\left|#1\right|}		
\newcommand{\norm}[1]{\left\lVert#1\right\rVert}		
\newcommand{\mat}[1]{\begin{bmatrix} #1 \end{bmatrix}}     

\newcommand{\mT}{\ \text{mT}}      				
\newcommand{\mV}{\ \text{mV}}				
\newcommand{\pc}{\text{\%}}					
\renewcommand{\deg}{{}^{\circ}}				
\newcommand{\um}{\ \upmu\text{m}}		
\newcommand{\us}{\ \upmu\text{s}}		
\newcommand{\ns}{\ \text{ns}}			

\newcommand{\gin}[1]{g_\mathrm{in#1}} 
\newcommand{\gout}{g_\mathrm{out}}
\newcommand{\tin}{t_\mathrm{ramp\, in}}
\newcommand{\tout}{t_\mathrm{ramp\, out}}
\newcommand{\tso}{t_\mathrm{SO}}
\newcommand{\nso}[1]{n^{#1}_\mathrm{SO}}
\newcommand{\zee}[2]{\varDelta^{#1}_{#2}}

\newcommand{\supmat}{SI} 
\newcommand{\SupRefFiggs}[1]{\figref{supfig:gs}}
\newcommand{\SupRefFigcharge}[1]{\figref{supfig:charge}}
\newcommand{\SupRefFiggcorr}[1]{\figref{supfig:g_corr}}

\begin{document}

\title{Spatial uniformity of g-tensor and spin-orbit interaction in germanium hole spin qubits}

\newcommand{\zrl}{IBM Research Europe -- Zurich, Säumerstrasse 4, 8803 Rüschlikon, Switzerland}
\newcommand{\yorktown}{IBM Research, T.\,J.\,Watson Research Center, 1101 Kitchawan Road, Yorktown Heights, New York 10598, USA}

\author{Inga~\surname{Seidler}}
\email{inga.seidler@ibm.com}
\affiliation{\zrl}

\author{Bence~\surname{Hetényi}}
\affiliation{\zrl}

\author{Lisa~\surname{Sommer}}
\affiliation{\zrl}

\author{Leonardo~\surname{Massai}}
\affiliation{\zrl}

\author{Konstantinos~\surname{Tsoukalas}}
\affiliation{\zrl}

\author{Eoin~G.~\surname{Kelly}}
\affiliation{\zrl}

\author{Alexei~\surname{Orekhov}}
\affiliation{\zrl}

\author{Michele~\surname{Aldeghi}}
\affiliation{\zrl}

\author{Stephen~W.~\surname{Bedell}}
\affiliation{\yorktown}
\author{Stephan \surname{Paredes}}
\affiliation{\zrl}

\author{Felix~J.~\surname{Schupp}}
\affiliation{\zrl}
\author{Matthias~\surname{Mergenthaler}}
\affiliation{\zrl}

\author{Gian~\surname{Salis}}
\affiliation{\zrl}
\author{Andreas~\surname{Fuhrer}}
\affiliation{\zrl}
\author{Patrick~\surname{Harvey-Collard}}
\email{phc@zurich.ibm.com}
\affiliation{\zrl}

\begin{abstract}
Holes in Ge/SiGe heterostructures are now a leading platform for semiconductor spin qubits, thanks to the high confinement quality, two-dimensional arrays, high tunability, and larger gate structure dimensions. One limiting factor for the operation of large arrays of qubits is the considerable variation in qubit frequencies or properties resulting from the strongly anisotropic \gtens{}. 
We study the \gtenss{} of six and seven qubits in an array with a Y geometry across two devices. We report a mean distribution of the tilts of the \gtens{}'s out-of-plane principal axis of around $1.1 \deg$, where nearby quantum dots are more likely to have a similar tilt. Independently of this tilt, and unlike simple theoretical predictions, we find a strong in-plane \gtens{} anisotropy with strong correlations between neighboring quantum dots.
Additionally, in one device where the principal axes of all g-tensors are aligned along the [100] crystal direction, we extract the spin-flip tunneling vector from adjacent dot pairs and find a pattern that is consistent with a uniform Dresselhaus-like spin-orbit field. 
The Y arrangement of the gate layout and quantum dots allows us to rule out local factors like electrostatic confinement shape or local strain as the origin of the preferential direction.
Our results reveal  long-range correlations in the spin-orbit interaction and \gtenss{} that were not previously predicted or observed, and could prove critical to reliably understand \gtenss{} in germanium quantum dots.
\end{abstract}

\date{3 October 2025}

\maketitle

\section{Introduction}\label{sec1}

Semiconductor hole spin qubits in \GeSiGe{} heterostructures provide large two-dimensional quantum dot (QD) arrays \cite{borsoi_shared_2024}, high confinement quality \cite{stehouwer_exploiting_2025}, high tunability \cite{riggelen2021}, and larger gate structure dimensions. 
High fidelity qubit operations \cite{lawrie_simultaneous_2023,wang_operating_2024} have been demonstrated utilizing all-electrical manipulation due to the strong spin-orbit interaction which provides a drive mechanism \cite{hendrickx2020a}, enables operational sweet-spots for qubit coherence \cite{hendrickx_sweet-spot_2024} and determines the qubit frequencies. The inability to predict and engineer the strongly anisotropic spin \gtens{} \cite{watzinger_heavy-hole_2016} is limiting for the operation of large numbers of qubits.  The large variation of effective \gfacts{} generates a large spread in qubit frequencies \cite{wang_operating_2024}. Measured values of the effective \gfacts{} vary in the range of 4 to 16 for out-of-plane magnetic fields and 0.06 to 0.62 for in-plane applied magnetic fields \cite{hendrickx_fast_2020,miller_effective_2022,jirovec_singlet-triplet_2021,john_two-dimensional_2024,tsoukalas_dressed_2025,zhou2025_High_fidelity}. 
While several experiments study fixed magnetic field directions, a full map of the three-dimensional (3D) \gtens{} is rarely known, in particular for planar germanium heterostructures \cite{hendrickx_sweet-spot_2024}. Meanwhile, the coherence sweet-spots that suppress the nuclear spin interaction \cite{hendrickx_sweet-spot_2024}, the Rabi frequency and charge-noise sensitivity \cite{hendrickx_sweet-spot_2024}, the two-qubit \gfact{} differences and the singlet-triplet gap \cite{jirovec_singlet-triplet_2021,kelly_identifying_2025} all depend quite sensitively on the precise tilt and shape of the \gtens{}.

The strong out-of-plane anisotropy of the \gtens{} is a trait of the heavy-hole character of the states. The stronger QD confinement in the growth direction than the in-plane directions pins the largest component of the \gtens{} to the growth direction \cite{watzinger_heavy-hole_2016}. The strength of the anisotropy can vary depending on the heavy-hole/light-hole gap, which depends on the intrinsic strain, and the degree of heavy- and light-hole orbital mixing \cite{wang_modeling_2024,sammak_shallow_2019,scappucci_germanium_2021}. Small tilts from the growth direction (out-of-plane) in experiments \cite{hendrickx_sweet-spot_2024} can be attributed to strain \cite{abadillo-uriel_hole-spin_2023}.

In addition to the large out-of-plane anisotropy, a small in-plane anisotropy is theoretically predicted for elongated confinement potentials mixing in higher orbitals \cite{brickson_using_2024,bosco_squeezed_2021}, with an implied dependence on gate voltages. For single holes, the amount of anisotropy predicted is relatively small, around 10 to $50\pc$ \cite{martinez_variability_2025,sarkar2025}. Experiments have also reported voltage tunability of the effective \gfact{} \cite{john_two-dimensional_2024,hendrickx_sweet-spot_2024}. However, the observed in-plane anisotropy is orders of magnitude stronger than the theoretical prediction \cite{hendrickx_sweet-spot_2024,brickson_using_2024,martinez_variability_2025}. Furthermore, the numerical values of the in-plane \gfact{} are often quite far from predictions. Theoretical studies of uniaxial in-plane strain show a larger range of \gfact{} modulation than what is achievable with gate voltages \cite{mauro_strain_2024,martinez_hole_2022}. However, strains of this magnitude can only be achieved with device engineering and are much larger than reported values of gate induced strain for planar heterostructures \cite{mauro_strain_2024,corley-wiciak_nanoscale_2023}. 

Here, we measure the full 3D \gtenss{} for two devices with six and seven qubits each using a vector magnet. We observe a distribution of out-of-plane tilts of the principal axes, and a strong correlation between the measured in-plane \gtenss{}. The device layout with seven QDs arranged in a Y geometry is key to differentiating layout symmetries from other effects as the origin of the observed spatial correlations. We consider and exclude the effects of a common sample tilt on the observations, and dot-local effects such as gate induced strain or gate induced confinement asymmetry. Additionally, we probe the avoided crossing of the singlet and spin-polarized triplet states of adjacent QDs, which is used for initialization and readout of the qubits. Using the knowledge of the \gtenss{} and of the tunneling momentum vector imposed by the Y device layout, we extract the spin-flip tunneling term induced by spin-orbit interaction. Surprisingly, the observed pattern is most consistent with the presence of a global uniform Dresselhaus-like spin-orbit mechanism. We discuss the possible physical mechanisms and implications for large arrays.

\section{Results}\label{sec2}

\begin{figure}[tbp]
    \centering
    \includegraphics[width=\columnwidth]{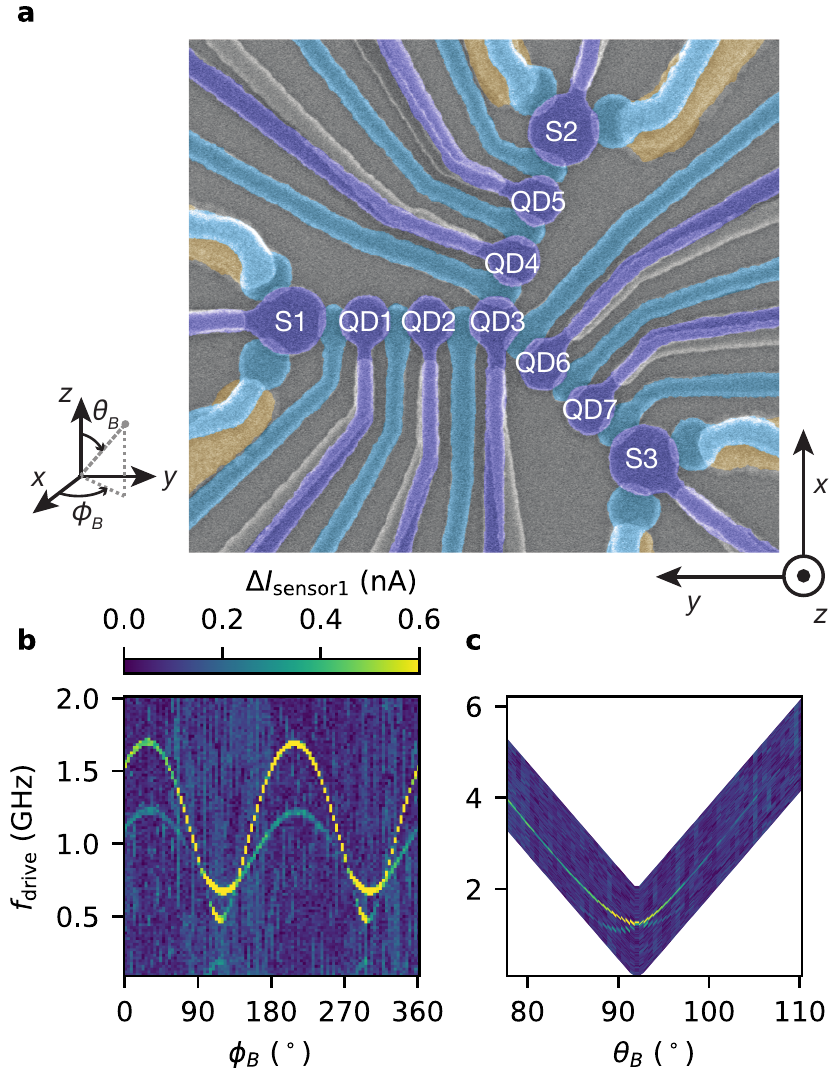}
    \caption{Device geometry and \gfact{} measurements.  
    \figletter{a}~Colored scanning electron micrograph of a Y junction device. The QDs used for qubit formation are labeled QD$i$, and those used as charge sensors are labeled S$i$. 
    \figletter{b-c}~Qubit frequencies of QD1 and QD2 for varying in-plane magnetic field angle $\phi_B$ (b) and out-of-plane angle $\theta_B$ (c). The sweeps are performed at a constant magnetic field strength of $100 \mT$.}
    \label{fig:device}
\end{figure}

We form six (device A) or seven (device B) single hole spin qubits confined in a strained \GeSiGe{} heterostructure quantum well, similar to Refs.\ \cite{tsoukalas_dressed_2025,massai2024_impact}. The holes are electrostatically confined underneath individual plunger gates in an identical device as \figref{fig:device}a. Barriers between neighboring QDs control their tunnel couplings. The charge state of the QDs is measured using the three larger sensing dots S1-S3 operated in transport. The device is tuned to the single hole regime for QD1-QD7 (see Supplementary Information (\supmat{}) for charge stability diagrams \secref{sec:device_tuning}). All plungers and barriers are virtualized with respect to the charge sensors and the QD occupations, so that the QDs can be independently tuned. The dot layout consists of three identical arms with angles $[112.5\deg,112.5\deg,135\deg]$ between them. The deviation from a $120\deg$ rotational symmetry accommodates space for the plunger of the central QD (QD3). To initialize and read out the spin states, we operate each arm of the device as a double QD (DQD) system and perform latched Pauli spin blockade (PSB) \cite{harvey-collard_high-fidelity_2018,kelly_identifying_2025}.

The Larmor frequency of each of the qubits is mapped as a  function of the direction of the applied external magnetic field in \figref{fig:device}b-c. For each measurement, a DQD in one device arm is prepared in the $\ket{\downarrow\downarrow}$ spin state. We initialize via an adiabatic ramp from a \Stwozero{} to a $\ket{{\downarrow\downarrow}(1,1)}$ state. The adiabatic ramp rate is optimized only for a specific magnetic field direction. However, an imperfect initialization of the $\ket{\downarrow\downarrow}$ state only decreases the visibility of the qubit frequency in our measurement and is therefore not a problem. 
We apply a qubit ac drive to one of the DQD plungers. The lever arm of the plungers is sufficient to drive either of the qubits. As we determine the spin state via PSB using the inverse of the adiabatic initialization ramp, all spin states except $\ket{\downarrow\downarrow}$ are blockaded and the readout is insensitive to which spin was flipped \cite{kelly_identifying_2025}. For high fidelity qubit operations, the initialization ramp would need to be calibrated for each specific magnetic field direction. 

The qubit frequency is especially sensitive to small remanent magnetic fields perpendicular to the sample plane. These can appear as asymmetries in the \gfact{} polar plots. We neutralize the out-of-plane remanent field to less than $0.5 \mT$ with the help of a spin funnel measurement \cite{petta_coherent_2005} following a procedure described in the \supmat{} \secref{sec:Spin_funnel}. To further minimize the effect of small offsets in the magnet fields, we measure \gfacts{} at a field strength of $100 \mT$.

\begin{figure*}[tbp]
    \centering
    \includegraphics[width=\textwidth]{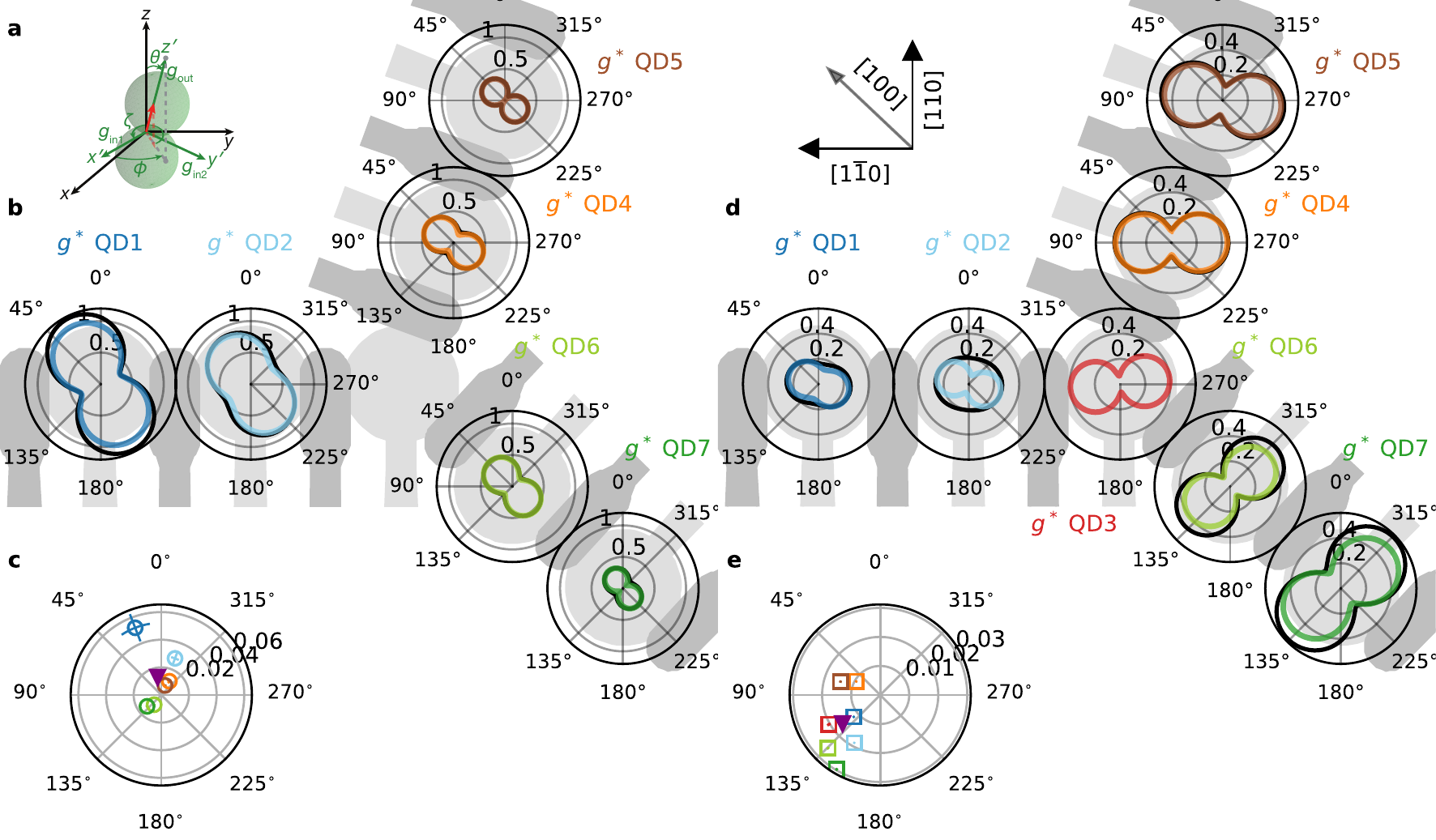}
    \caption{\gtens{} orientations.  
    \figletter{a}~Schematic of a \gtens{}, with the lab frame in black, the principal axis frame in green, and a unit vector $\uvct{z}'$ along the $z'$ axis representing the tilt (red). The $zyz$ Euler rotation angles $\zeta$, $\theta$ and $\phi$ describe the transformation between the two frames. 
    \figletter{b}~In-plane \gtens{} cross-sections of \geff{} in the lab frame (black) and in the individual tilt frame of each \gtens{} (color) for device A. The individual tilt frames are tilted such that the new $z$ axis aligns with every individual \gtens{}'s $z'$ axis without including additional rotations (\supmat{} \secref{sec:tilt}). The cross-section plots are arranged according to the device layout.  See \supmat{} \figref{supfig:gs_A} and \figref{supfig:gs_B} for numerical values.
    \figletter{c}~Tilt visualization for device A by projecting of the $\uvct{z}'$ unit vectors on the $x{-}y$ plane with matching colors. The solid inverted triangle represents the projection of the average tilt. 
    \figletter{c-d}~In-plane \geff{} (d) and tilts (e) for device B.}
    \label{fig:g_eigen_y}
\end{figure*}

The effective \gfact{} \geff{} is extracted from data like \figref{fig:device}b-c by dividing the measured qubit frequency $f$ by the magnitude of the applied magnetic field $B$: $\geff \muB B = h f$, where $\muB$ is the Bohr magneton. 
The general \gtens{} is modeled in the lab frame as a real symmetric $3\times3$ matrix $\tens{g}_\text{lab} = \mtrx{R} \operatorname{diag}(\gin1, \gin2, \gout) \mtrx{R}^{-1}$, illustrated in \figref{fig:g_eigen_y}a, with principal axis values $\gin1 < \gin2 \ll \gout$ and rotation matrix $\mtrx{R}=\mtrx{R}(\zeta,\theta,\phi)$ describing the extrinsic rotations around the axes $zyz$ (see \supmat{} \secref{sec:g_fit}) \cite{hendrickx_sweet-spot_2024}. 
The full \gtens{} fit requires, in addition to the in-plane measured qubit frequencies, two distinct out-of-plane directions. Detailed fit data are presented in \supmat{} \figrefs{supfig:dqd12}{supfig:dqd67}. The \gtenss{} of both devices are represented in \figref{fig:g_eigen_y}b,d, where the black line represents the data in the lab frame.

Strikingly, the in-plane \geff{} in the lab frame features a strong anisotropy and similarity between nearest neighbors. In device A, the long axis aligns approximately along a $45\deg$ angle (\figref{fig:g_eigen_y}b, black line) that corresponds to a [100] crystal direction. Such large anisotropy is not predicted or observed in any published work to date \cite{martinez_variability_2025,scappucci_germanium_2021}. Theoretically, a sample tilt could generate this common direction in the $x{-}y$ plane of the lab frame. The different magnitudes of $\gout \approx 11.0 \pm 0.5 \approx 50 \gin1$ and the small tilt angles $\theta \lesssim 3\deg$ make it difficult to represent the tilt of the \gtenss{} on a polar plot. Instead, we consider a unit vector $\uvct{z}'$ along the principal axis $z'$ of the \gtens{} (\figref{fig:g_eigen_y}a). The projection of this unit vector onto the $x{-}y$ plane shows that there is a relatively large spread of tilts  (\figref{fig:g_eigen_y}c,e). A sample tilt can only add a common tilt for all \gtenss{} and would therefore cluster the projections around a specific point. A best guess for this sample tilt would be the average of the $\uvct{z}'$ vectors (solid purple triangle). Here, the spread of the $\uvct{z}'$ projections is larger than (device A) or comparable to (device B) the average tilt, and much larger than the predicted spread $\lesssim 0.2\deg$ from electrostatic disorder alone \cite{martinez_variability_2025}. We note that the $95\pc$  uncertainty, indicated by bars, is well below the spread of the data.

To address the question of the impact of sample tilt on lab frame observations, we consider each \gtens{} in its individual tilt frame. We define this frame such that the new $z$ axis aligns with the individual \gtens{}'s $z'$ axis using a rotation away from the $z$ axis by an angle $\theta$ in the direction of $z'$, $\mtrx{T}(\theta,\phi)$. A detailed discussion of the different frames is given in \supmat{} \secref{sec:tilt}.
The results are shown in \figref{fig:g_eigen_y}b,d as colored lines. 
In these frames, different for each dot, the minimal and maximal \geff{} on the $x_{\mathrm{tilt,}i}{-}y_{\mathrm{tilt},i}$ plane match the principal axis values $\gin1$ and $\gin2$ of the \gtens{}. The anisotropy of the in-plane components can be enhanced (QD7 A, QD5 A) or reduced (QD1 A); however, it always holds that $g^*_{\mathrm{tilt},i} \leq g^*_\mathrm{lab}$, as in this individual tilt frame the 2D polar plot shows \geff{} along the waist of the 3D \gtens{} surface. 
As we show in the \supmat{} \secref{sec:tilt}, this also holds when visualizing the \gtenss{} in a common average (sample) tilt frame. Therefore, we have ruled out that the  appearance of a correlated directionality between adjacent in-plane \gtens{} cuts could be attributed to sample tilt.

\begin{figure}[tbp]
    \centering
    \includegraphics[width=\columnwidth]{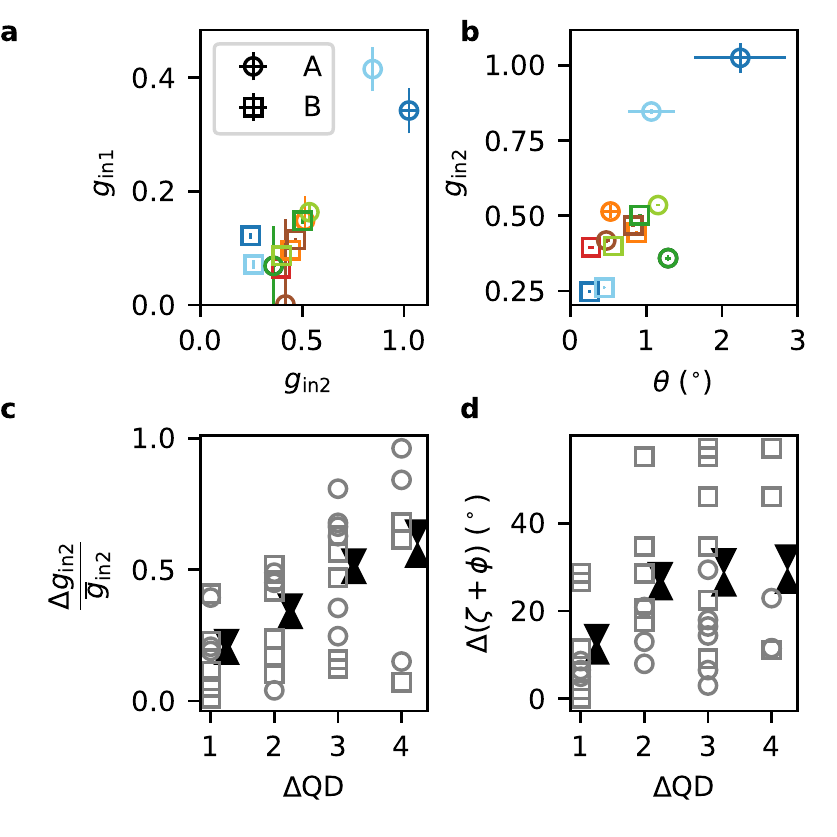}
    \caption{ Correlations between \gtenss{}.  
    \figletter{a}~Correlation between $\gin1$ and $\gin2$, revealing that they are likely to increase together.
    \figletter{b}~Correlation between the tilt angle $\theta$ and $\gin2$, revealing that larger tilt angles are associated with larger $\gin2$. Here, $\theta$ is calculated in the average tilt frame (i.e., not the lab frame).
    \figletter{c-d}~Relative change in $\gin2$ (c) and absolute change in orientation angle $\zeta + \phi$ ((d); \eqnref{eq:oriangle}) as a function of distance in units of QD pitch. The mean ($\blacktriangle$) and root-mean-square ($\blacktriangledown$) values are plotted with a slight offset alongside, revealing a trend that closer QDs are more likely to have similar $\gin2$ and in-plane orientation $\zeta + \phi$.
    }
    \label{fig:g_corr_main}
\end{figure}

We quantify the correlations between the different \gtenss{} of both devices in \figref{fig:g_corr_main}. Noticeably, a larger $\gin1$ often coincides with a larger $\gin2$ (\figref{fig:g_corr_main}a). 
Larger tilt angles $\theta$ correlate with larger $\gin2$ (and $\gin1$) (\figref{fig:g_corr_main}b). Here, $\theta$ is calculated in the average tilt frame (i.e., not the lab frame). These correlations indicate a possible common underlying mechanism between the tilt angle and the renormalization of $\gin1,\gin2$.
In \figref{fig:g_corr_main}c, we plot the relative change $\Delta\gin2/\bar{g}_\mathrm{in2} = 2 \abs{{\gin2}_{,i}-{\gin2}_{,j}}/({\gin2}_{,i}+{\gin2}_{,j})$ as a function of the distance $\Delta\mathrm{QD}$ between QDs $i$ and $j$ along the Y array, in units of QD pitch, for all possible pairs. We do similarly for the angle that parametrizes the direction of the in-plane anisotropy, $\zeta+\phi$ (see \eqnref{eq:oriangle}), and plot the direction angle difference in \figref{fig:g_corr_main}d. This analysis reveals that the spread (root-mean-square) of the observed differences increases as the QDs get further apart in the array. That such a trend is visible is a strong indicator that the microscopic mechanism at play has a correlation length comparable to the size of the Y array.

The \gtens{} is strongly influenced by the underlying spin-orbit mechanisms. The unique device geometry suggest another opportunity to investigate spin-orbit interaction in these devices, this time considering tunneling between pairs of QDs. Specifically, we study the dynamics of the \STm{} anticrossing in device A, which is also of general relevance during initialization and readout of the individual qubits \cite{kelly_identifying_2025}. The \STm{} gap involves a contribution from (perpendicular) \gfact{} differences, as well as a spin-flip tunneling transition from a \Stwozero{} to a \Tmoneone{} state \cite{jirovec_dynamics_2022}. This tunneling transition has its momentum vector oriented along the DQD axis \cite{nichol_quenching_2015}, therefore giving us access to three independent and known momentum directions along the device's Y geometry. 

\begin{figure*}[tbp]
    \centering
    \includegraphics[width=\textwidth]{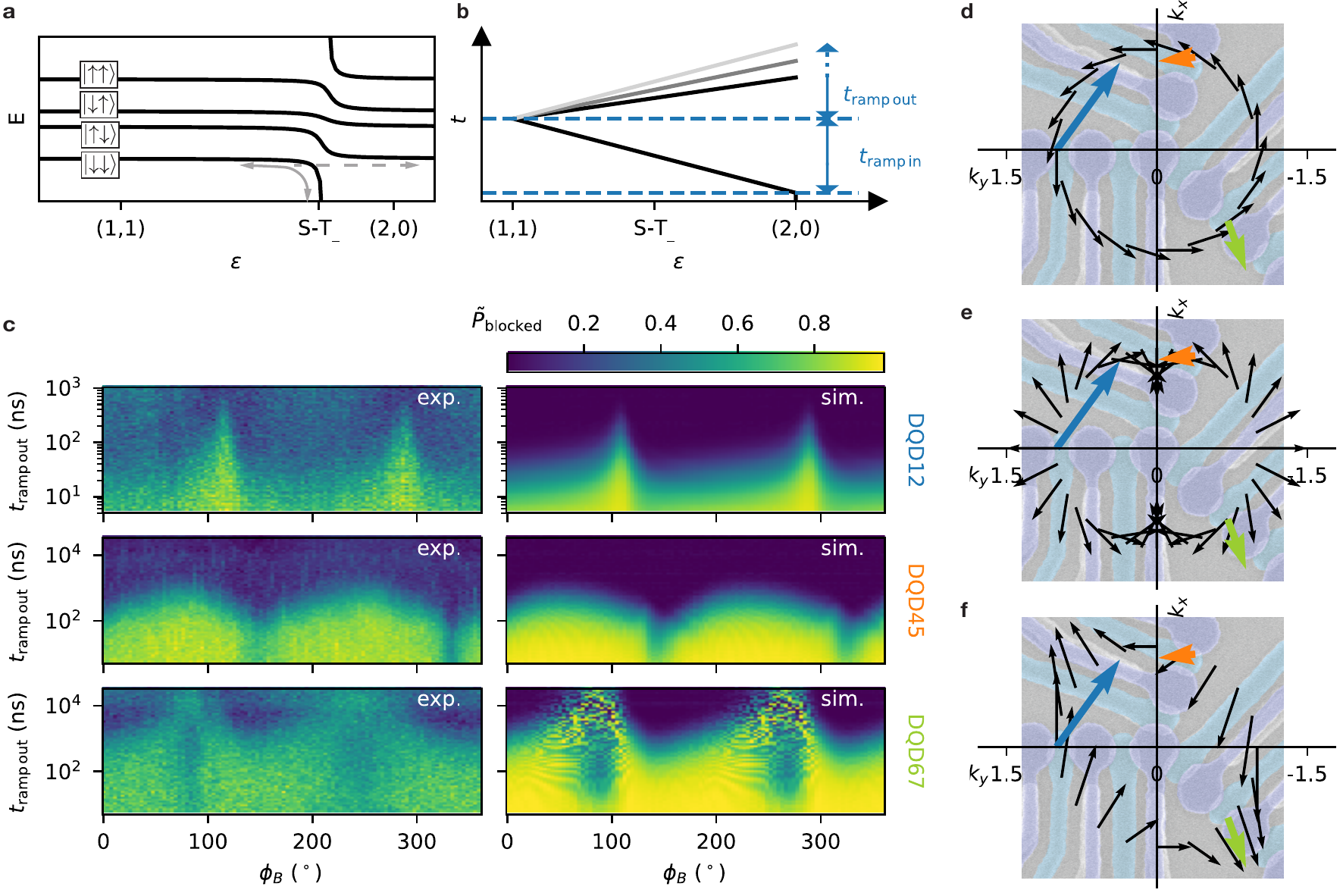}
    \caption{Spin-orbit fields in device A. 
    \figletter{a}~Energy diagram of two spins in a DQD. The \STm{} anti-crossing with size $2\Delta_{ST}$ is  probed with detuning ramps indicated with the grey arrows. 
    \figletter{b}~Schematic of the applied pulse sequence for fixed $\tin$ and variable $\tout$. 
    \figletter{c}~Measured and simulated magnetic field angle dependence of the return probability of a blocked spin state for a tunneling between QD1 and QD2 with $\tin=1\us$  (below: QD5 and QD4, QD7 and QD6, with $\tin=30\us$)  and variable $\tout$ with an external magnetic field of $7\mT$ applied in-plane. 
    \figletter{d-f}~Fitted spin-orbit vectors in momentum space assuming a momentum aligned with the design DQD axis (color according to the referenced DQD). For comparison a linear Rashba-like (d), cubic Rashba-like (e) and Dresselhaus-like (f) spin-orbit field is plotted (black).}
    \label{fig:ST}
\end{figure*}

To measure the spin-orbit vector $\vct n_\mathrm{SO}$ \cite{geyer_anisotropic_2024,kelly_identifying_2025}, in device A, we initialize a DQD in one of the array arms in a \Stwozero{} state (\figref{fig:ST}a), ramp the detuning ($\varepsilon$) with time $\tin$ to the $(1,1)$ charge state, ramp back to the $(2,0)$ charge state with time $\tout$ and perform latched PSB readout (\figref{fig:ST}b) \cite{nichol_quenching_2015,harvey-collard2019}. 
Reliably achieving an adiabatic ramp is difficult due to a strong magnetic field dependence (\figref{fig:ST}c) and a relatively small interdot tunnel coupling, which results in incoherent mixing at long ramp times. Therefore, $\tin$ is chosen such that adiabaticity is given for most field angles. 

To extract the \STm{} gap, a Landau-Zener approximation is often used to fit the measured probability data \cite{nichol_quenching_2015,harvey-collard2019,jirovec_dynamics_2022}. However, the approximation breaks down when the gap is comparable to the qubit energy splitting \cite{kelly_identifying_2025}, which is the case for specific magnetic field angles due to the in-plane anisotropy of the \gtenss{}. In addition, the adiabaticity of the ramp to $(1,1)$ is not satisfied for all magnetic field angles. Therefore, we instead model the data with a time-dependent five state Hamiltonian and calculate the time-evolution of an initial state for the two sequential ramp experiments (for details see \supmat{} \secref{sec:SO}). Knowledge of the full 3D \gtenss{} and DQD tunnel couplings are used as fixed input parameters while the parameters of the spin-orbit interaction are fitted to obtain the spin-orbit field $\vct n_\mathrm{SO}$. For all three arms, the experimental probability data can be accurately modeled (\figref{fig:ST}c). 

The $\vct n_\mathrm{SO}$ vectors can be arranged according to the tunneling momentum of each DQD (\figref{fig:ST}d-f). 
The remarkable uniformity in the \gtens{} orientations in device A suggests that perhaps the spin-orbit field is also somewhat uniform. For Ge holes, the usual symmetries are of the direct Rashba type, $\hat H_\text{R} = \alpha (\hat k_y \hat\sigma_x - \hat k_x \hat\sigma_y)$, and cubic Rashba, $\hat H_\text{R3} = \alpha_3 (\hat k_y^3 \hat\sigma_x - \hat k_x^3 \hat\sigma_y)$.
While three effective spin-orbit field vectors are insufficient to fit the coefficients of the different possible spin-orbit mechanisms, a qualitative comparison based on symmetry can be made. The measured spin-orbit field does not match a linear Rashba-like or cubic Rashba-like spin-orbit field (\figref{fig:ST}d,e) which have previously been observed in \GeSiGe{} heterostructures \cite{moriya_cubic_2014,mizokuchi_hole_2017} and are normally considered dominant \cite{bosco_squeezed_2021,terrazos2021_theory}. Instead, a Dresselhaus-like symmetry $\hat H_\text{D} = \beta (\hat k_x \hat\sigma_x - \hat k_y \hat\sigma_y)$ is most consistent with the measured spin-orbit field (\figref{fig:ST}f). While a Dresselhaus spin-orbit field is not predicted from the crystal symmetry of the \GeSiGe{} heterostructure, symmetry breaking due to the interface of the quantum well or strain within the quantum well could be the origin of this spin-orbit mechanism \cite{rodriguez-mena_linear--momentum_2023} as previously observed for Si-MOS spin qubit devices \cite{jock_silicon_2018}. Remarkably, Dresselhaus-like spin-orbit interaction shows the same symmetry as the in-plane cross-sections of the \gtenss{}.

\section{Discussion}

Our results reveal the existence of a long-range correlation between dot \gtenss{}. 
Theoretical calculations of \gtenss{} predict an influence of the confinement potential and strain profile due to the gate layout of the device \cite{abadillo-uriel_hole-spin_2023,brickson_using_2024}. In our devices, both the electrostatic confinement and local strain induced by the gate structure would introduce a rotational symmetry of approximately $120\deg$ to the in-plane \gtens{} cross-sections, which is not observed. However, they can still contribute small variations in uniformity (\supmat{} \secref{sec:tuna}). 
Electrostatic disorder (i.e., charge traps) should not produce correlations beyond nearest neighbors, if any.
If strain is the origin of the correlated orientation of the \gtens{} cross-sections, a global strain field on the length scale of the QD array must be present. One possible source could be strain introduced by the growth, which is visible on the device surface as a crosshatch pattern and approximately matches the required length scale of $1\um\times1\um$ \cite{stehouwer_germanium_2023}. 
We also remark that the sample tilt might be quite different from the average tilt, if the mechanism causing the tilt has a similar directionality as the one causing the in-plane anisotropy.

\section{Conclusion}

In summary, we have studied the \gtenss{} of two identical six- and seven-qubit devices with QDs arranged in a Y geometry. 
Our results reveal that the \gtenss{} have relatively large tilts of 1 to $2\deg$ with respect to the growth direction of the heterostructure. Though these tilts may seem small, the effects are dramatically amplified by the large out-of-plane \gfact{} and are much larger than predictions for electrostatic disorder or elongated orbitals.
The in-plane \gfact{} also displays a very strong anisotropy with obvious correlations between neighboring QDs that can extend far into the array. One device even has all six QDs aligning with the [100] crystal direction. Such a pattern is not consistent with local strain or electrostatic confinement potential as the origin of the anisotropy. Furthermore, the effective spin-orbit field, independently probed with a spin-flip tunneling experiment where the Y geometry allows to generate momenta along three different known directions and reveals the presence of a Dresselhaus-like spin-orbit symmetry, the same directionality as the g-tensors. Finally, the correlations between the tilt angle and $\gin1, \gin2$ hints at a common mechanism linking tilt, in-plane anisotropy strength and long range effects.

Our measurements reveal unexpected properties of the \gtens{} and spin-orbit interaction in planar Ge, and could help understand and ultimately engineer the \gtens{}.  In-plane magnetic field orientations are specifically of interest as they allow for all-electrical qubit driving at MHz frequencies when operating in a low field regime and hyperfine noise suppression. Initialization, readout, qubit frequencies, driving speed and decoherence sweet spots all depend sensitively on this anisotropy. Operating many qubits simultaneously with a common magnetic field direction  could unlock the full potential of hole spin qubits in Ge, and understanding their \gtens{} is the key to this goal.

\begin{acknowledgments}
The authors thank Michael Stiefel and all the Cleanroom Operations Team of the Binnig and Rohrer Nanotechnology Center (BRNC) for their help and support.
\paragraph*{\bf Funding}
This research was funded in part by {NCCR SPIN}, a National Centre of Competence in Research, funded by the Swiss National Science Foundation (grants  \mbox{51NF40-180604} and \mbox{51NF40-225153}) and by the Swiss National Science Foundation (grant \mbox{200021-188752}).
\paragraph*{\bf Author contributions}
I.S.\ performed the experiments and data analysis.
B.H.\ and P.H.C.\ developed the spin-orbit interaction model.
L.M., E.G.K., G.S., P.H.C.\ and I.S.\ developed part of the measurement software.
F.J.S.\ and M.M.\ fabricated the device, and K.T.\ designed the gate layout with inputs from P.H.C.\ and A.F. 
A.O.\ and M.A.\ contributed to the interpretation.
L.S., I.S., K.T., A.O., M.A.\ and P.H.C.\ developed parts of the device functionality.
S.W.B.\ grew the heterostructures. 
S.P., L.S., A.F.\ and P.H.C.\ contributed to the development of the experimental setup. 
I.S.\ and P.H.C.\ wrote the manuscript with input from all authors. 
P.H.C.\ supervised the project with help from G.S.\ and A.F.
\paragraph*{\bf Competing interests}
The authors declare no competing interests.
\end{acknowledgments}

\bibliographystyle{apsrev4-1-title} 
\bibliography{references.bib}


\makeatletter
\renewcommand\thesection{\mbox{S\arabic{section}}}
\makeatother
\setcounter{section}{0}     

\newcounter{supfigure} \setcounter{supfigure}{0} 
\makeatletter
\renewcommand\thefigure{\mbox{S\arabic{supfigure}}}
\makeatother

\newcounter{suptable} \setcounter{suptable}{0} 
\makeatletter
\renewcommand\thetable{\mbox{S\arabic{suptable}}}
\makeatother

\newenvironment{supfigure}[1][]{\begin{figure}[#1]\addtocounter{supfigure}{1}}{\end{figure}\ignorespacesafterend}
\newenvironment{supfigure*}[1][]{\begin{figure*}[#1]\addtocounter{supfigure}{1}}{\end{figure*}\ignorespacesafterend}
\newenvironment{suptable}[1][]{\begin{table}[#1]\addtocounter{suptable}{1}}{\end{table}\ignorespacesafterend}
\newenvironment{suptable*}[1][]{\begin{table*}[#1]\addtocounter{suptable}{1}}{\end{table*}\ignorespacesafterend}
\newenvironment{suplongtable}[1][]{\addtocounter{suptable}{1}\begin{longtable}[#1]}{\end{longtable}\ignorespacesafterend}

\makeatletter
\renewcommand\theequation{S\arabic{equation}}
\renewcommand\theHequation{S\arabic{equation}} 
\makeatother
\setcounter{equation}{0}

\newcommand{\RefMainFigOne}{Fig.\,\ref{fig1}} 
\newcommand{\RefMainFigTwo}{Fig.\,\ref{fig:g_eigen_y}} 
\newcommand{\RefMainFigThree}{Fig.\,\ref{fig3}} 

\onecolumngrid 
\clearpage
\newpage
{\centering
\large\textbf
{Supplementary information for: \\ Spatial uniformity of g-tensor and spin-orbit interaction in germanium hole spin qubits} \\ \rule{0pt}{12pt}
}

\section{Device tuning}
\label{sec:device_tuning}

Device A is tuned to the single hole regime for the seven QDs (\figref{supfig:charge}). Virtual gates (linear combination of physical gate voltages) are used to control the QDs. To suppress exchange interactions with the spin in the central QD (QD3), the barrier voltages controlling the tunnel couplings to QD3 are increased by $100 \mV$ to compared to the tuning point of the charge stability diagrams. During the \gtens{} measurements, all QDs are occupied by a single hole. 
\begin{supfigure*}[htbp]
    \centering
    \includegraphics[width=\textwidth]{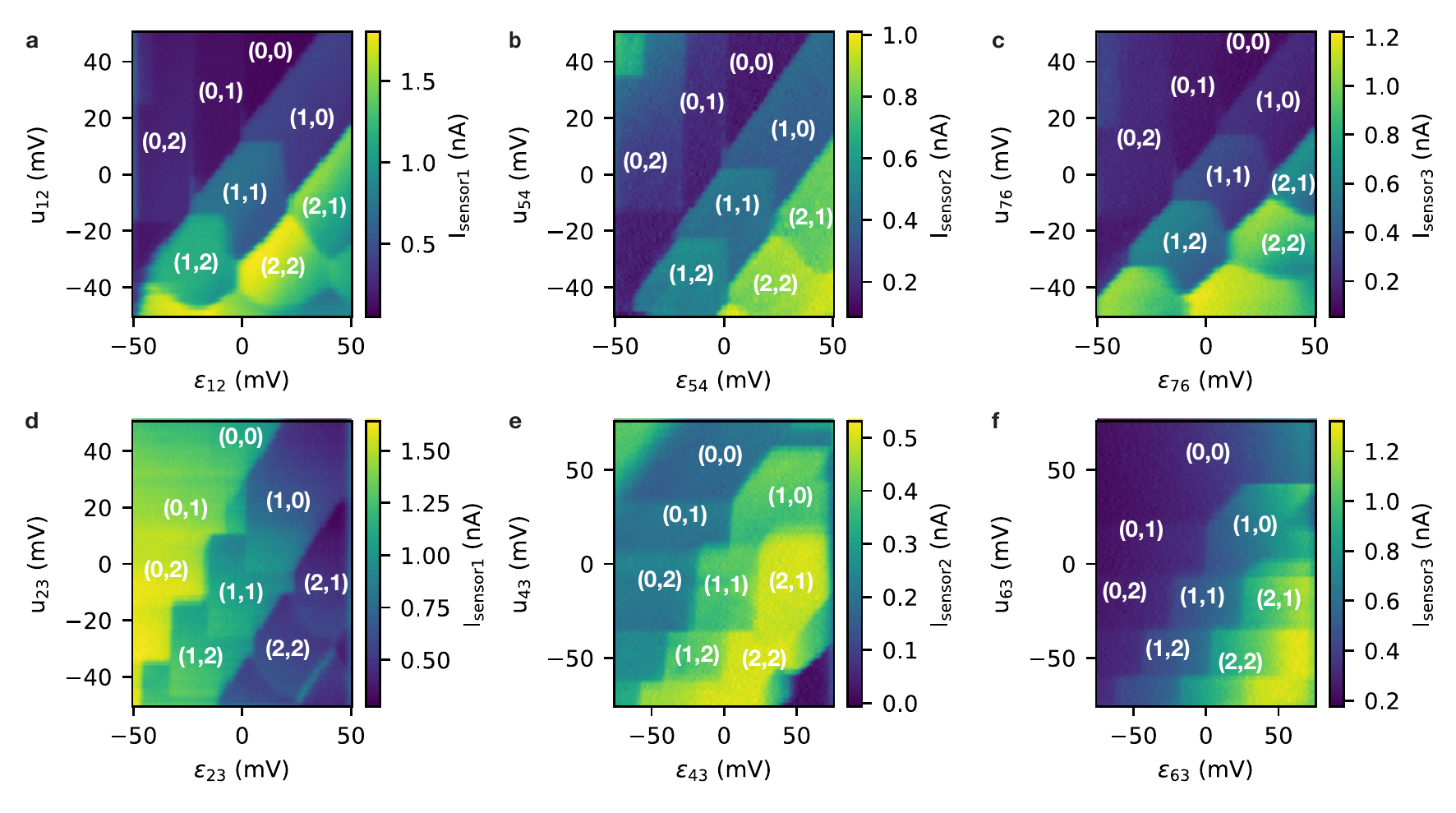}
    \caption{ Charge stability diagrams for the outer (a-c) and inner (d-f) DQDs with the charge occupation marked. The occupation label $(n_i,n_j)$ corresponds to axis labels $\varepsilon_{ij}$ and $u_{ij}$ for DQD $i,j$. The inner loop of the measurement is the detuning voltage sweep ($\varepsilon_{ij}$), which is swept from negative to positive voltages for measurements a-c and reversed for measurements d-f.}
    \label{supfig:charge}
\end{supfigure*}

For device B, the hole configuration differs. Each device arm is tuned to the $(0,2)$ charge configuration with the two holes in the QD towards the center. Central dot QD3 is unoccupied except during the measurement of its \gtens{}. For the measurement of each \gtens{}, the corresponding DQD is detuned to the $(1,1)$ charge configuration, while the rest of the device remains in the steady state configuration. To measure the \gtens{} of QD3, the spin state is detected via latched PSB between QD3 and QD4. We do not observe a strong influence of the hole configuration or the gate voltage on the \gtens{} (see \secref{sec:tuna}).

\section{Remanent magnetic field calibration}
\label{sec:Spin_funnel}

The vector magnet shows a hysteretic behavior, a normal behavior for these types of superconducting magnets. After applying fields above $\sim 50\mT$ on one axis, we observe up to $3\mT$ of remanent field when this axis is nominally set back to $0\mT$.  Due to the large out-of-plane $g_\text{out} \approx 11$, the qubit measurements near in-plane fields are highly sensitive to remanent out-of-plane magnetic fields ($B_z$). For example, if $B_z = 1 \mT$, the equivalent in-plane field (in qubit frequency) is approximately $B_z g_\text{out} / g_\text{in} \approx 110 \mT$ (assuming $g_\text{in} \approx 0.1$). Remanent fields lead to \gfact{} polar plots that are not symmetric around the origin.

We determine the remanent field using a spin funnel measurement as in \figref{supfig:funnel}a. If an out-of-plane field remains, the spin funnel does not open fully at $B_x=0$, as shown in \figref{supfig:funnel}a. For a minimized remanent $B_z$, the spin funnel opens at $B_x=0$, as in \figref{supfig:funnel}b. The magnitude of the remanent $B_z$ field can be determined by stepping $-2\mT < B_z < 2\mT$ in small increments and finding the funnel with the largest opening. 

In this work, the remanent field is zeroed out to less than $0.5 \mT$ by applying a larger opposite field in the coil, and then confirming the smallness of the remanent field by repeating the funnel measurements before sensitive measurements.
\begin{supfigure*}[tbp]
    \centering
    \includegraphics[width=14cm]{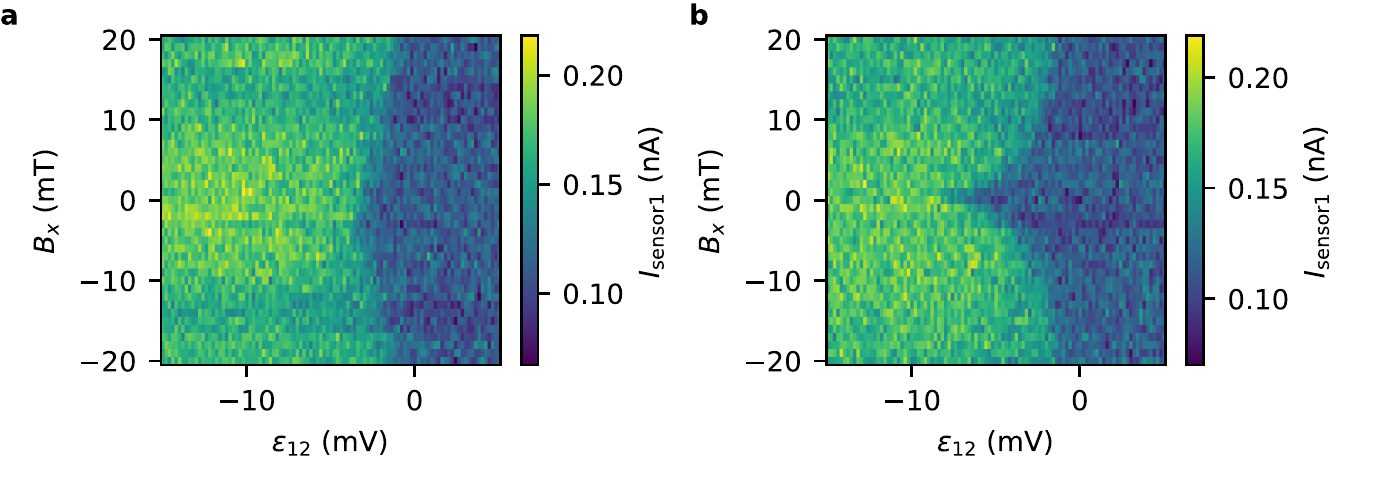}
    \caption{Spin funnels for calibration of remanent magnetic fields. \figletter{a}~Spin funnel with nominally $B_z=0$, but with remanent field of $\approx 0.5\mT$ out-of-plane. At $B_x=0\mT$, the funnel does not open.  \figletter{b}~Spin funnel with nominally $B_z=0\mT$ and a remnant out-of-plane field $\leq 0.5 \mT$. At $B_x=0\mT$, the funnel opens.}
    \label{supfig:funnel}
\end{supfigure*}
\FloatBarrier 

\begin{supfigure*}[tbp]
    \centering
    \includegraphics[width=\textwidth]{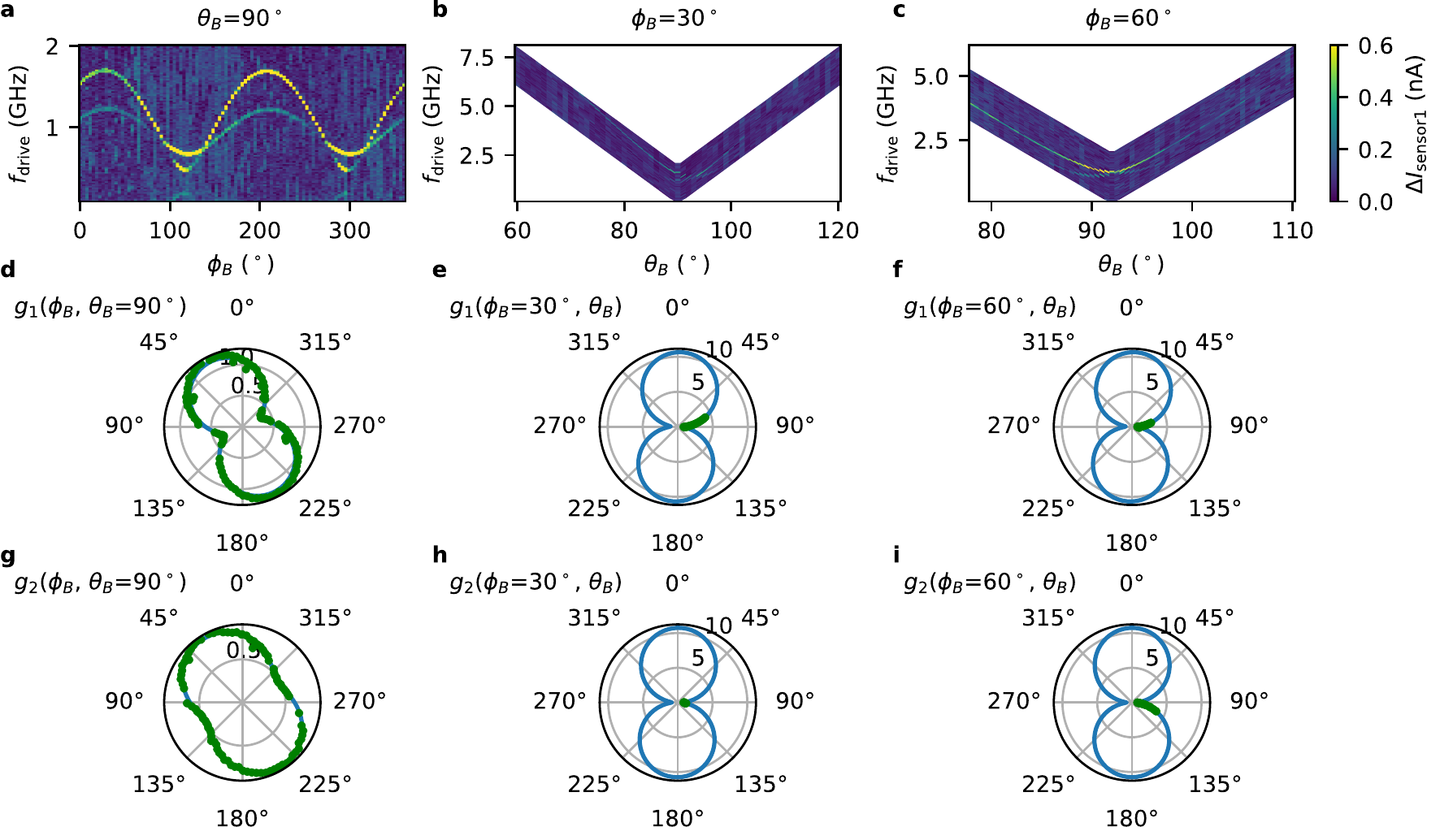}
    \caption{ Measurement of the hole \gtens{} of QD1 and QD2 (device A). \figletter{a-c} Qubit frequencies of QD1 and QD2 for magnetic field sweeps in-plane (a) and two out-of-plane directions (b,c) of the lab frame.  \figletter{d-f, g-i} Cross section of the \gtens{} of QD1 (d-f) and QD2 (g-i) in planes of the qubit frequency measurements of the lab frame. Dots indicate measurements of $\geff$ and the solid line corresponds to the fit of the \gtens{}.}
    \label{supfig:dqd12}
\end{supfigure*}
\begin{supfigure*}[tbp]
    \centering
    \includegraphics[width=\textwidth]{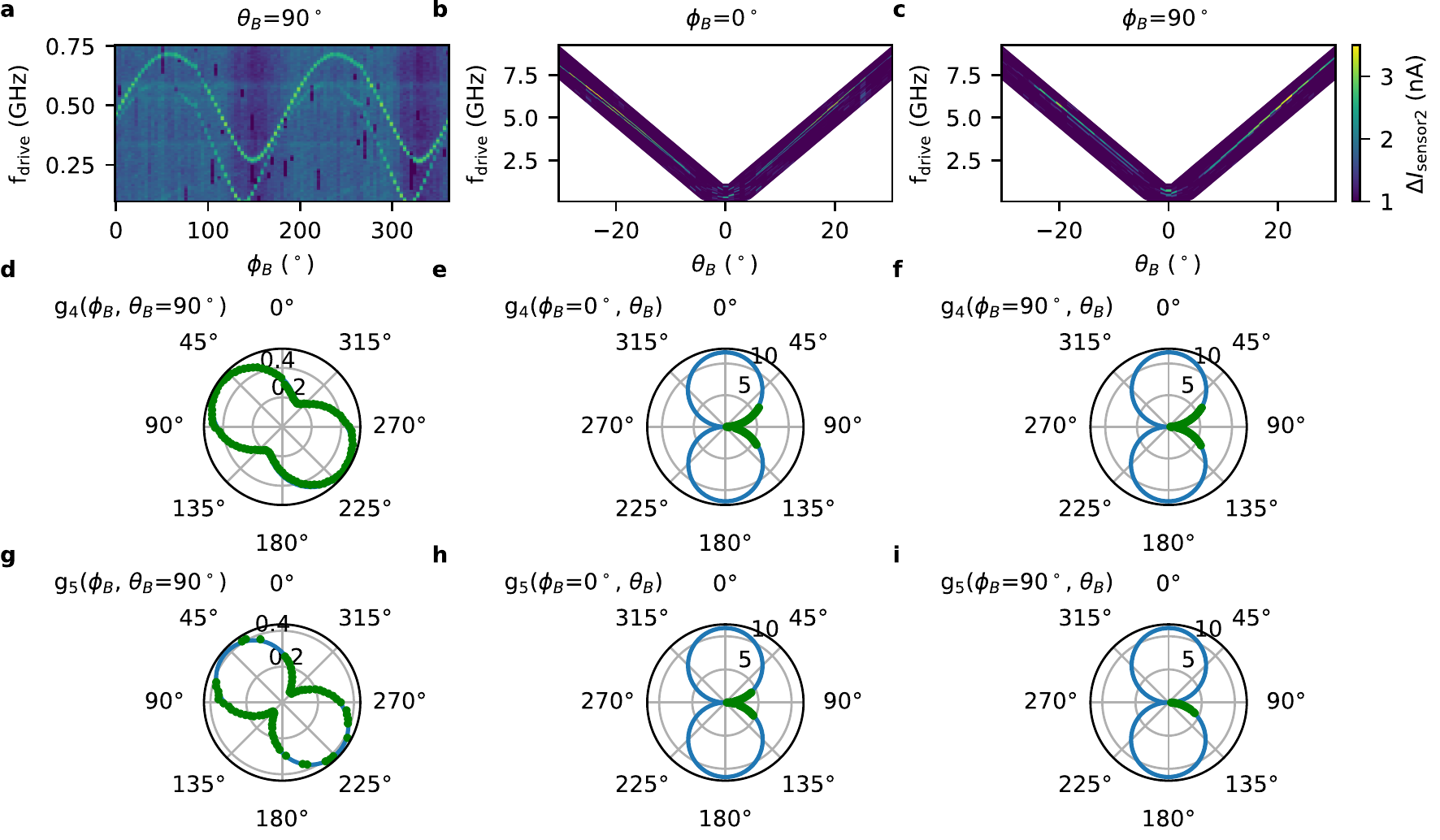}
    \caption{ Measurement of the hole \gtens{} of QD4 and QD5 (device A). For detailed description cf.\ \figref{supfig:dqd12}}
    \label{supfig:dqd45}
\end{supfigure*}
\begin{supfigure*}[tbp]
    \centering
    \includegraphics[width=\textwidth]{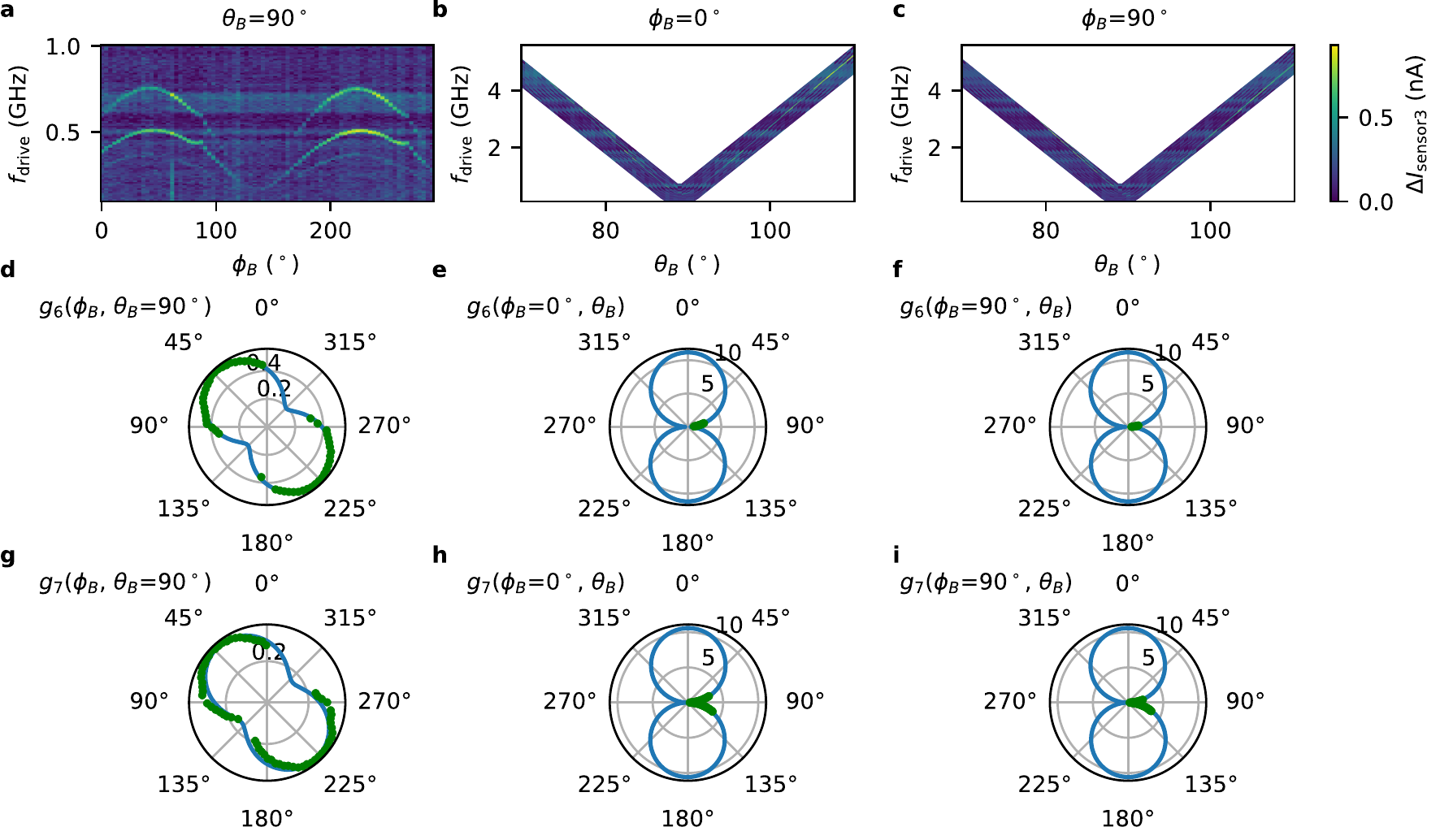}
    \caption{ Measurement of the hole \gtens{} of QD6 and QD7 (device A). For detailed description cf.\ \figref{supfig:dqd12}}
    \label{supfig:dqd67}
\end{supfigure*}

\FloatBarrier 

\section{\gtenspdf{} fitting}
\label{sec:g_fit}

The qubit frequencies are assigned to the different dot positions by comparing the different drive strengths of the QD plungers. The plunger corresponding to a qubit's position is not necessarily the most efficient to drive this qubit \cite{john_two-dimensional_2024}. To differentiate them, the plunger of the central QD is used, as it drives all six qubits peripherally and is a less efficient drive for the outer qubits in comparison to the inner qubits. 

To determine the \gtenss{}, the frequency of each qubit is tracked with respect to the applied magnetic field orientation. The magnetic field is varied in the lab ($\approx$ sample) plane and for two out-of-plane sweep directions, as in \figrefs{supfig:dqd12}{supfig:dqd67}. Three independent measurement planes are necessary to reconstruct the full \gtens{}. For each magnetic field $\vct B(\phi_B,\theta_B)$, where $B = \norm{\vct B}$ and $\uvct b = \vct B / B$, \geff{} is calculated from the qubit frequency using $g^{*} \muB B = h f$.

The general \gtens{} in the lab frame is modeled as a rotated diagonal $3\times3$ matrix with principal axis values $\gin1$, $\gin2$, $\gout$ and rotation matrix $\mtrx{R}(\zeta,\theta,\phi)$. The Euler rotation angles $\zeta$, $\theta$ and $\phi$ describe the extrinsic rotations around the axes $zyz$ \cite{hendrickx_sweet-spot_2024}:
\begin{align}
    \tens{g}_\mathrm{lab}=
    \mtrx{R}(\zeta,\theta,\phi) \mat{ \gin1 &0 &  0 \\  0 &  \gin2 & 0 \\  0& 0 & \gout } \mtrx{R}^{-1}(\zeta,\theta,\phi) ,
\end{align}
with
\begin{align}
    \mtrx{R}(\zeta,\theta,\phi) = 
    \mtrx{R}_z(\phi) \mtrx{R}_y(\theta) \mtrx{R}_z(\zeta) = 
    \mat{ 
    c_\phi c_\theta c_\zeta-s_\phi s_\zeta & -s_\phi c_\zeta-c_\phi c_\theta s_\zeta & c_\phi s_\theta \\
    s_\phi  c_\theta  c_\zeta+c_\phi s_\zeta& c_\phi c_\zeta-s_\phi  c_\theta s_\zeta& s_\phi s_{\theta}\\
    -s_{\theta} c_{\zeta}&                  s_{\theta} s_{\zeta}                     & c_{\theta}
    } ,
\end{align}
where $c_\phi=\cos{\phi}$, $c_\theta=\cos{\theta}$, $c_\zeta=\cos{\zeta}$, $s_\phi=\sin{\phi}$, $s_\theta=\sin{\theta}$ and $s_\zeta=\sin{\zeta}$. A ``lab'' index denotes the lab frame, a ``g'' index denotes the principal axis frame, and so on.

To determine the six principal axis values and rotation angles of $\tens{g}_\mathrm{lab}$, the experimental lab frame values of $\geff$ are fitted  with
\begin{align}
    g^*\left(\phi_B,\theta_B\right)
    &=
    \norm{\tens{g}_\mathrm{lab} \uvct{b}} 
    = 
    \norm{\tens{g}_\mathrm{lab}\left(\zeta,\theta,\phi,\gin1,\gin2,\gout\right)      \mat{ \sin{\theta_B}\cos{\phi_B} \\ \sin{\theta_B}\sin{\phi_B} \\  \cos{\theta_B}}} .
\end{align}
The raw data and fits are shown in \figrefs{supfig:dqd12}{supfig:dqd67} for device A.

The extracted \gtens{} parameters are depicted in \figref{supfig:gs_A} and \figref{supfig:gs_B} for device A and B, respectively.

\begin{supfigure*}[tbp]
    \centering
    \includegraphics[width=\textwidth]{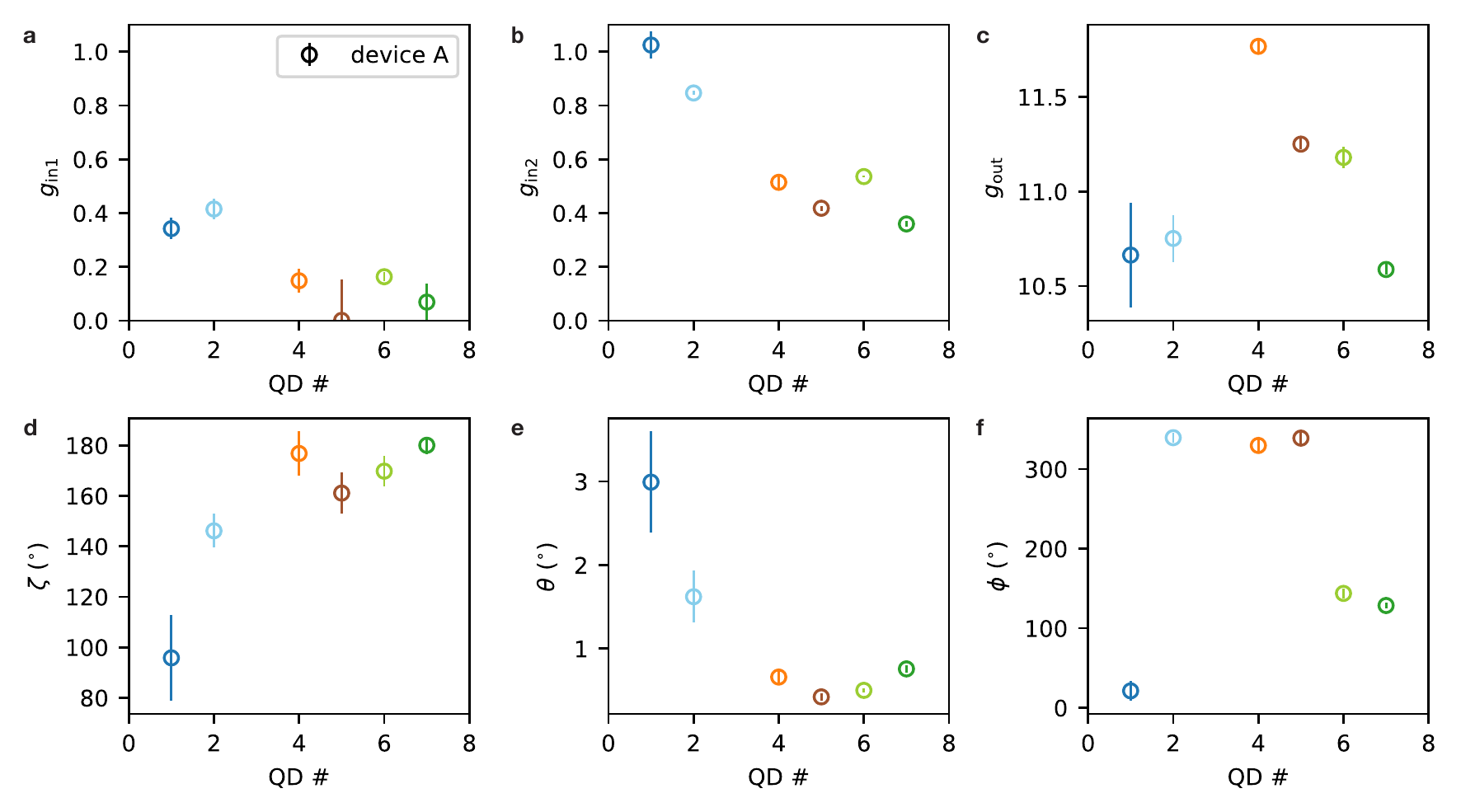}
    \caption{ Fit results for the six \gtenss{} of device A. \figletter{a-c} Fitted principal \gtens{} values. \figletter{d-f} Fitted $zyz$ Euler rotation angles $\phi$, $\theta$, $\zeta$. Polar plots shown in \figref{fig:g_eigen_y}.}
    \label{supfig:gs_A}
\end{supfigure*}
\begin{supfigure*}[tbp]
    \centering
    \includegraphics[width=\textwidth]{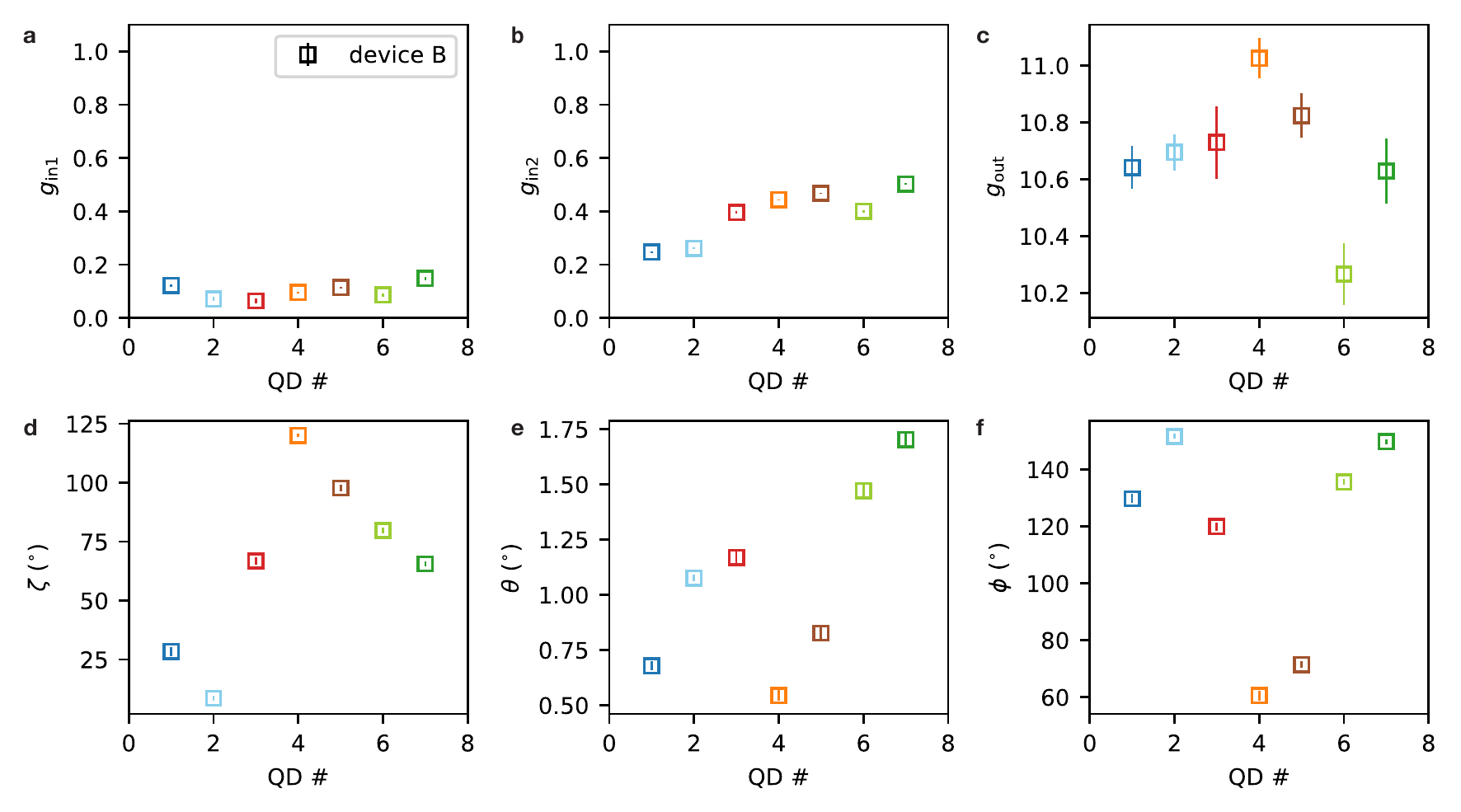}
    \caption{ Fit results for the seven \gtenss{} of device B. \figletter{a-c} Fitted principal \gtens{} values. \figletter{d-f} Fitted $zyz$ Euler rotation angles $\phi$, $\theta$, $\zeta$. Polar plots shown in \figref{fig:g_eigen_y}.}
    \label{supfig:gs_B}
\end{supfigure*}

\section{Sample tilt evaluation and impact}
\label{sec:tilt}

\begin{supfigure*}[tbp]
    \centering
    \includegraphics[width=\textwidth]{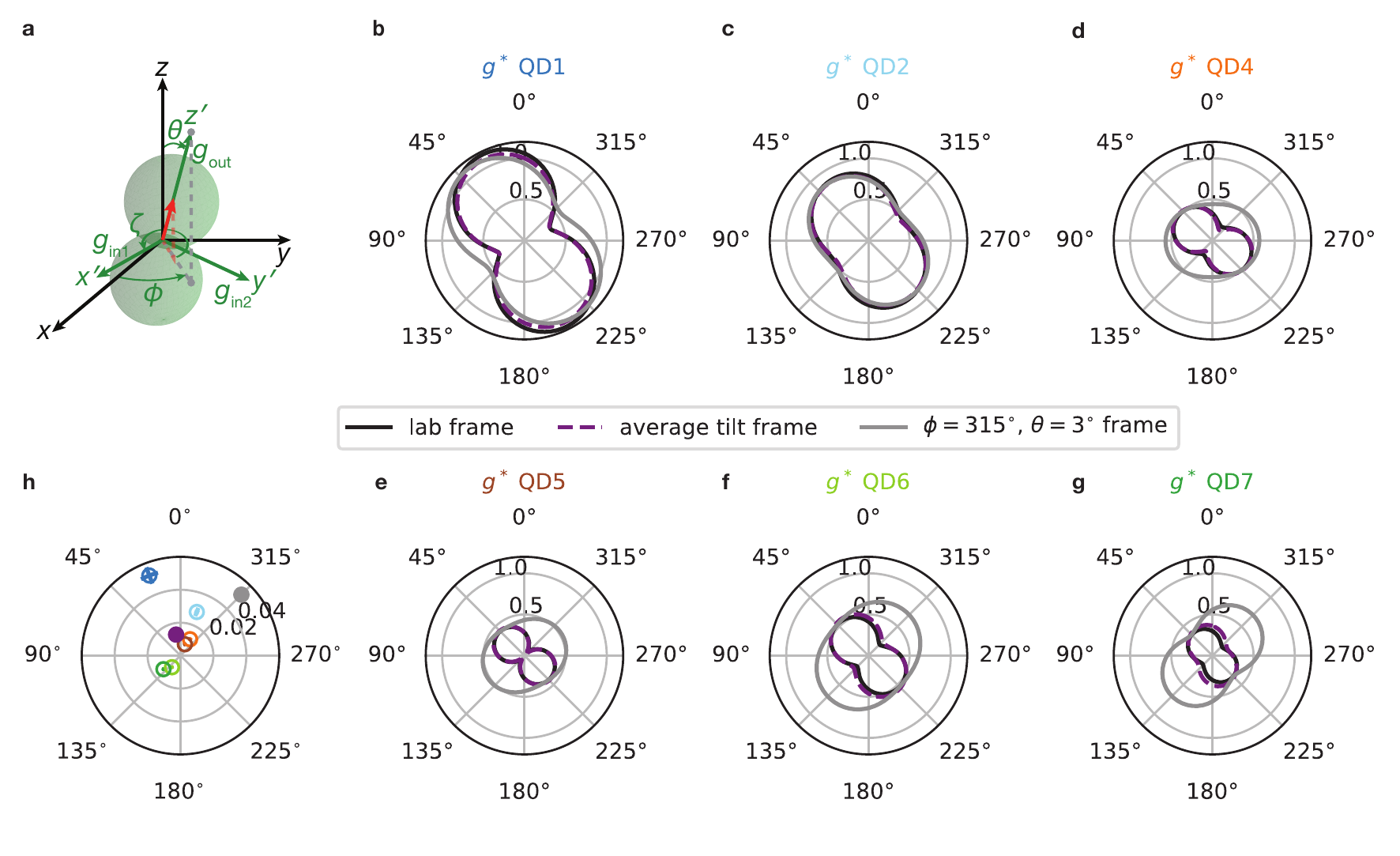}
    \caption{Device A \gtenss{} in the average tilt frame. \figletter{a} Schematic of a \gtens{} with the lab frame (black), its eigenframe (green) and a unit vector along the $z'$ axis (red) including the $zyz$ Euler rotation angles $\zeta$, $\theta$ and $\phi$ that describe the transformation between the two frames.  \figletter{b-g} Effective \geff{} in the $x{-}y$ plane of the lab frame, average tilt frame and a frame tilted by $3\deg$ in the direction $\phi=315\deg$ for each QD. All \geff{} are based on the fits of the individual \gtenss{}. The average tilt and random tilt frame projection are added to panel h for comparison. \figletter{h} Tilt of $\uvct{z}'$ of each \gtens{} represented by the projection of the unit vectors on the $x{-}y$ plane. The magnitude of \geff{} along the $z'$ direction is not taken into account. The color of each projection matches the title color of the corresponding \gtens{} in panels b-g.}
    \label{supfig:g_tilt}
\end{supfigure*}

As discussed in the main text, the tilt of a \gtens{} ``peanut'' shape can be visualized by the projection of a unit vector along the $z'$ axis of the \gtens{} on the $x{-}y$ plane (\figref{supfig:g_tilt}a). 
To calculate this projection, we first consider the unit vector along the $z'$ axis, which in the principal axis frame of each \gtens{} is represented by the unit vector $\uvct{z}'_\mathrm{g} = \mat{0 & 0 & 1}^\mathrm{T}$.
Following the \gtens{} description of \secref{sec:g_fit}, this unit vector represented in the lab frame is given by
\begin{equation}
    \uvct{z}'_\mathrm{lab} 
    = 
    \mtrx{R}(\zeta,\theta,\phi)  \uvct{z}'_\mathrm{g}
    =
    \mat{ \cos\phi\,\sin\theta \\ \sin\phi\,\sin\theta \\ \cos\theta }
\end{equation}
The tilt can therefore be visualized as a point on the $x{-}y$ plane with angle $\phi$ and radius $\sin{\theta}$, similar to pencils in a pencil holder viewed from the top.

A sample tilt (in the colloquial sense) can only add a common tilt $\mtrx{T}$ (a rotation matrix, defined below) to all \gtenss{} and would therefore cluster the projections around a specific angle and tilt strength $\theta$. 
The average tilt is our best guess at the sample tilt. However, we should note the possibility that the mechanism creating the tilt has a bias towards a certain direction, in which case the average tilt would not be a good proxy for the sample tilt even with enough data samples. The average tilt vector $\vct{\bar s}$ is calculated by averaging the $\uvct{z}'$ unit vectors and normalizing the result, 
\begin{equation}
    \vct{\bar{s}}_\mathrm{lab}
    =
    \dfrac{\sum_i\uvct{z}'_{i,\text{lab}}}{\lVert\sum_i\uvct{z}'_{i,\text{lab}}\rVert}
    =
    \dfrac{\sum_i \mtrx{R}(\zeta_i,\theta_i,\phi_i)  \uvct{z}'_g }{\lVert\sum_i \mtrx{R}(\zeta_i,\theta_i,\phi_i)  \uvct{z}'_g\rVert}  .
\end{equation} 

We define a tilt as a rotation away from the $z$ axis by an angle $\tilde{\theta}$ in the direction $\tilde{\phi}$. The rotation axis in the $x{-}y$ plane is given by $\uvct{a}=\mat{-\sin{\tilde{\phi}} , \cos{\tilde{\phi}} , 0}$, perpendicular to the direction of the tilt. 
The tilt matrix is given by
\begin{align}
    \mtrx{T}(\tilde{\theta},\tilde{\phi})
    &=
    \cos{\tilde{\theta}}\,\mtrx{I}+ \sin{\tilde{\theta}} \left[\uvct{a}\right]_\mathrm{\times}+\left(1-\cos{\tilde{\theta}}\right)\uvct{a}\otimes \uvct{a}\\
    &=
    \begin{bmatrix}
        \sin^2{\tilde{\phi}}\,(1-\cos{\tilde{\theta}})+\cos{\tilde{\theta}} &
        -\sin{\tilde{\phi}}\cos{\tilde{\phi}}\,(1-\cos{\tilde{\theta}})&
        \cos{\tilde{\phi}}\sin{\tilde{\theta}}\\
        -\sin{\tilde{\phi}}\cos{\tilde{\phi}}\,(1-\cos{\tilde{\theta}})&
        \cos^2{\tilde{\phi}}\,(1-\cos{\tilde{\theta}})+\cos{\tilde{\theta}}&
        \sin{\tilde{\phi}}\sin{\tilde{\theta}}\\
        -\cos{\tilde{\phi}}\sin{\tilde{\theta}}&
        -\sin{\tilde{\phi}}\sin{\tilde{\theta}}&
        \cos{\tilde{\theta}}\\
    \end{bmatrix} ,
\end{align}
where $\left[\uvct{a}\right]_\mathrm{\times}=\uvct{a}\times \mtrx{I}$ is the cross product matrix, $\uvct{a}\otimes \uvct{a}$ the outer product and $\mtrx{I}$ the identity matrix. 
The tilt matrix can alternatively by described using the Euler angles, noting that $\mtrx{T} = \mtrx{R}_z(\tilde\phi) \mtrx{R}_y(\tilde\theta) \mtrx{R}_z(-\tilde\phi)$. To find the average tilt frame, we determine $\bar{\theta}$ and $\bar{\phi}$ such that 
\ma{\vct{\bar s} = \mtrx{T}{\left(\bar{\theta},\bar{\phi}\right)}  \uvct{z},} where $\uvct{z}$ is the lab $z$ unit vector. This condition is satisfied for $\bar\theta=\arccos(\uvct{z}\cdot\vct{\bar s})$ and $\bar\phi=\mathrm{atan2}(\uvct{x}\cdot\vct{\bar s},\uvct{y}\cdot\vct{\bar s})$.
We can relate the tilt matrix to the Euler rotation matrix defined earlier noting that 
\begin{align}
    \mtrx{R}(\tilde\zeta,\tilde\theta,\tilde\phi) = \mtrx{T}(\tilde\theta,\tilde\phi) \mtrx{R}_z(\tilde\zeta+\tilde\phi)  ,
    \label{eq:oriangle}
\end{align}
i.e., that the Euler angles represent a rotation in the $x{-}y$ plane by an angle $\zeta+\phi$ (``orientation angle''), followed by a tilt by an angle $\theta$ in the direction of $\phi$.

To visualize the effect of a sample tilt (or any other tilt) on the \gfact{} angular measurement, the \gtens{} is rotated to the new frame using
\begin{align}
    \tens{g}_\mathrm{tilt}
    =
    \mtrx{T}^{-1}(\tilde{\theta},\tilde{\phi}) \tens{g}_\mathrm{lab}  \mtrx{T}(\tilde{\theta},\tilde{\phi}) .
\end{align}
In the average tilt frame, \geff{} in the $x_\mathrm{{tilt}}{-}y_\mathrm{{tilt}}$ plane polar plot shows only small variations in comparison to \geff{} in the $x{-}y$ plane of the lab frame (\figref{supfig:g_tilt}) or the principal axis frame (\figref{fig:g_eigen_y}).  In particular, the common directionality in device A and spatial correlation (devices A and B) is also present in the average tilt frame. 

Considering a fictious additional frame with a tilt of $\theta=3\deg$ and $\phi=315\deg$ highlights that the effective in-plane \gfacts{} will increase and have reduced anisotropy when tilting away from the principal axis frame (grey line in \figref{supfig:g_tilt}b-g). It is possible to generate a non-uniform orientation of the in-plane \geff{} in such a frame. However, the \geff{} will be significantly larger than the lab frame \geff{}, signifying that the cut plane is too tilted. As the six different \gtenss{} tilt in varying directions, it is not possible to find a common frame with higher anisotropy and smaller \geff{} than the average tilt frame.  Only if the considered frame is the eigenframe, or closer to the eigenframe than the lab frame,  are the in-plane \geff{} reduced.
The presence of a sample tilt cannot be differentiated from a systematic tilt with respect to the growth direction. However, we can definitively eliminate a sample tilt as the origin of the directionality of the in-plane \gtenss{}. 

For further analysis of the \gtens{} parameters, we consider all the \gtenss{} in the average tilt frame of their respective samples. 
The principal axis values are by construction independent of the frame. However, the angles describing the \gtens{} orientation change with the frame of reference. 
In the average tilt frame, the \gtenss{} are given by
\begin{align}
    \vct{\bar{g}}_\mathrm{tilt}
    &=    \mtrx{R}(\zeta_\mathrm{tilt},\theta_\mathrm{tilt},\phi_\mathrm{tilt}) \mat{ \gin1 &0 &  0 \\  0 &  \gin2 & 0 \\  0& 0 & \gout } \mtrx{R}^{-1}(\zeta_\mathrm{tilt},\theta_\mathrm{tilt},\phi_\mathrm{tilt} ) ,
\end{align}
with 
\begin{align}
    \mtrx{R}(\zeta_\mathrm{tilt},\theta_\mathrm{tilt},\phi_\mathrm{tilt}) =
    \mtrx{T}^{-1}(\bar{\theta},\bar{\phi}) \mtrx{R}(\zeta,\theta,\phi) ,
\end{align}
where $\theta$, $\zeta$ and $\phi$ refer to the angles in the lab frame, and $\bar\theta$ and $\bar\phi$ are the angles parametrizing the average frame (as described above). The angles in the tilt frame are therefore given by
\begin{align}
    \theta_\mathrm{tilt} &= \arccos\!\big(\uvct{z}\cdot \mtrx{T}(\bar{\theta},\bar{\phi}) \mtrx{R}(\zeta,\theta,\phi)\uvct{z}\big) ,\\
    \phi_\mathrm{tilt} &= \operatorname{atan2}\!\big(\uvct{x}\cdot \mtrx{T}(\bar{\theta},\bar{\phi}) \mtrx{R}(\zeta,\theta,\phi)\uvct{z},\uvct{y}\cdot \mtrx{T}(\bar{\theta},\bar{\phi}) \mtrx{R}(\zeta,\theta,\phi)\uvct{z}\big) ,\\
    \zeta_\mathrm{tilt} &= \operatorname{atan2}\!\big(-\uvct{z}\cdot \mtrx{T}(\bar{\theta},\bar{\phi}) \mtrx{R}(\zeta,\theta,\phi)\uvct{x},\uvct{z}\cdot \mtrx{T}(\bar{\theta},\bar{\phi}) \mtrx{R}(\zeta,\theta,\phi)\uvct{y}\big) .
\end{align}

\section{\gtenspdf{} tunability}
\label{sec:tuna}

Theoretical models predict a dependence of the \gtens{} on the shape of the confinement potential \cite{bosco_squeezed_2021,brickson_using_2024}. In particular, it is one of the mechanisms that can introduce an in-plane anisotropy. 
We determine the influence of the confinement potential on the \gtens{} parameters by measuring $\tens{g}_6$ and $\tens{g}_7$ in dependence of two barrier gate voltages (\figref{supfig:B67_tuna} and \figref{supfig:B36_tuna}). For each dataset, the applied dc voltages and the pulse sequence for initialization and readout are identical. The change in barrier gate voltage is added as an adiabatic pulse at the manipulation point before applying the qubit drive, such that we can measure the qubit frequency splitting in the deformed potential. Each data point is based on the fitting of three independent planes of magnetic field sweeps as described in \secref{sec:g_fit}.

The interdot barrier B67 between QD6 and QD7 can tune  $\gin{1}$ and $\gin{2}$ in a small range (\figref{supfig:B67_tuna}). Interestingly, we do not observe a reduction of the in-plane anisotropy, as $\gin{1}$ and $\gin{2}$  are both reduced with lower barrier voltage. We note that at $\mathrm{vB67}=0\mV$, we are in a regime of large exchange coupling between QD6 and QD7, such that the eigenstates start to hybridize.

A larger voltage range can be measured for the variation of the barrier B36 between QD3 and QD6. It  tunes $\tens{g}_6$ while $\tens{g}_7$ remains mostly unchanged, as is expected due to the larger distance between the QD and gate (\figref{supfig:B36_tuna}).  

\begin{supfigure*}[tbp]
    \centering
    \includegraphics[width=\textwidth]{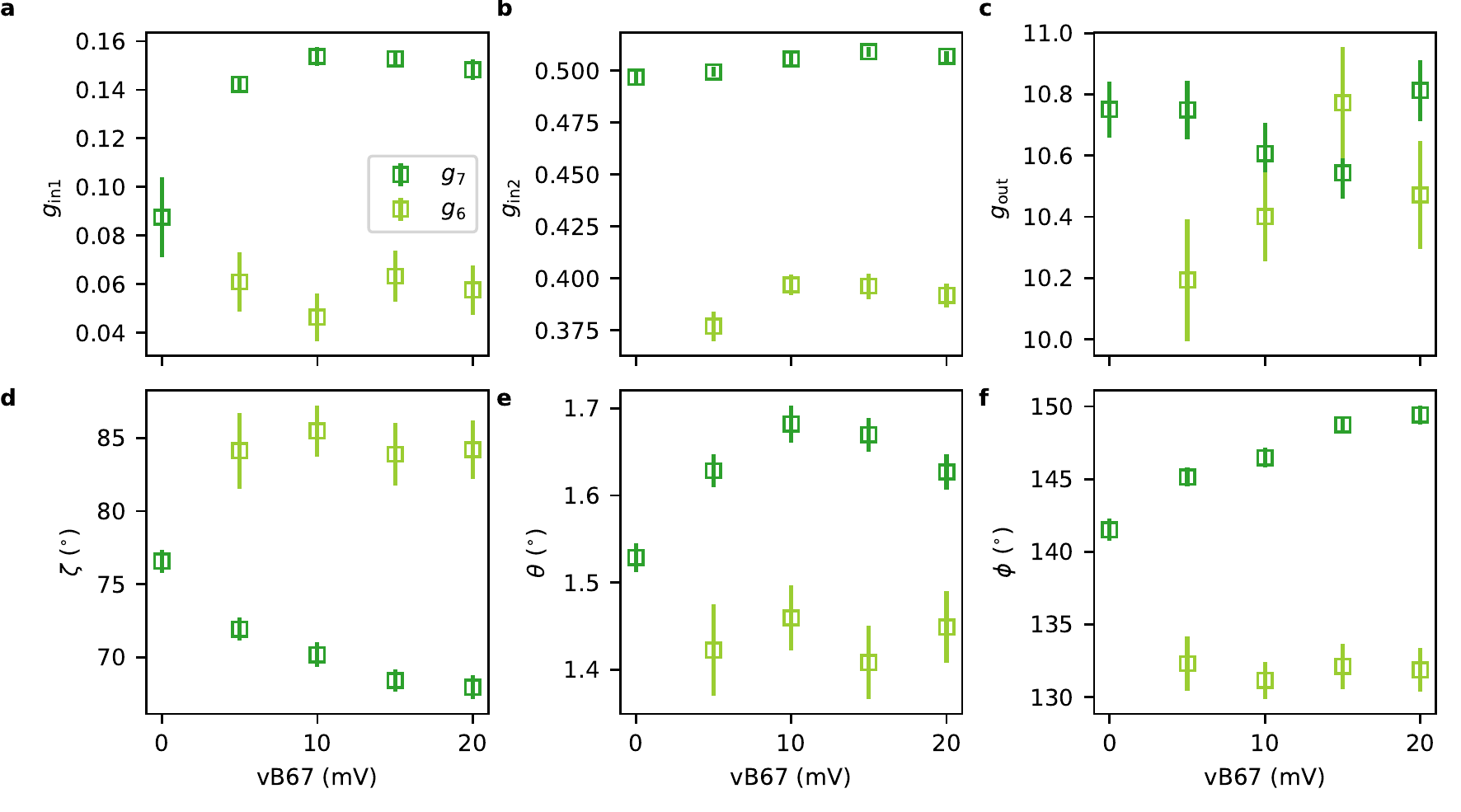}
    \caption{Gate tunability of $\tens{g}_6$ and $\tens{g}_7$ with interdot barrier B67 of device B. \figletter{a-f} Fitted principal axis \gtens{} values and $zyz$ Euler rotation angles for different barrier gate voltages vB67 applied to the barrier between QD6 and QD7. }
    \label{supfig:B67_tuna}
\end{supfigure*}

\begin{supfigure*}[tbp]
    \centering
    \includegraphics[width=\textwidth]{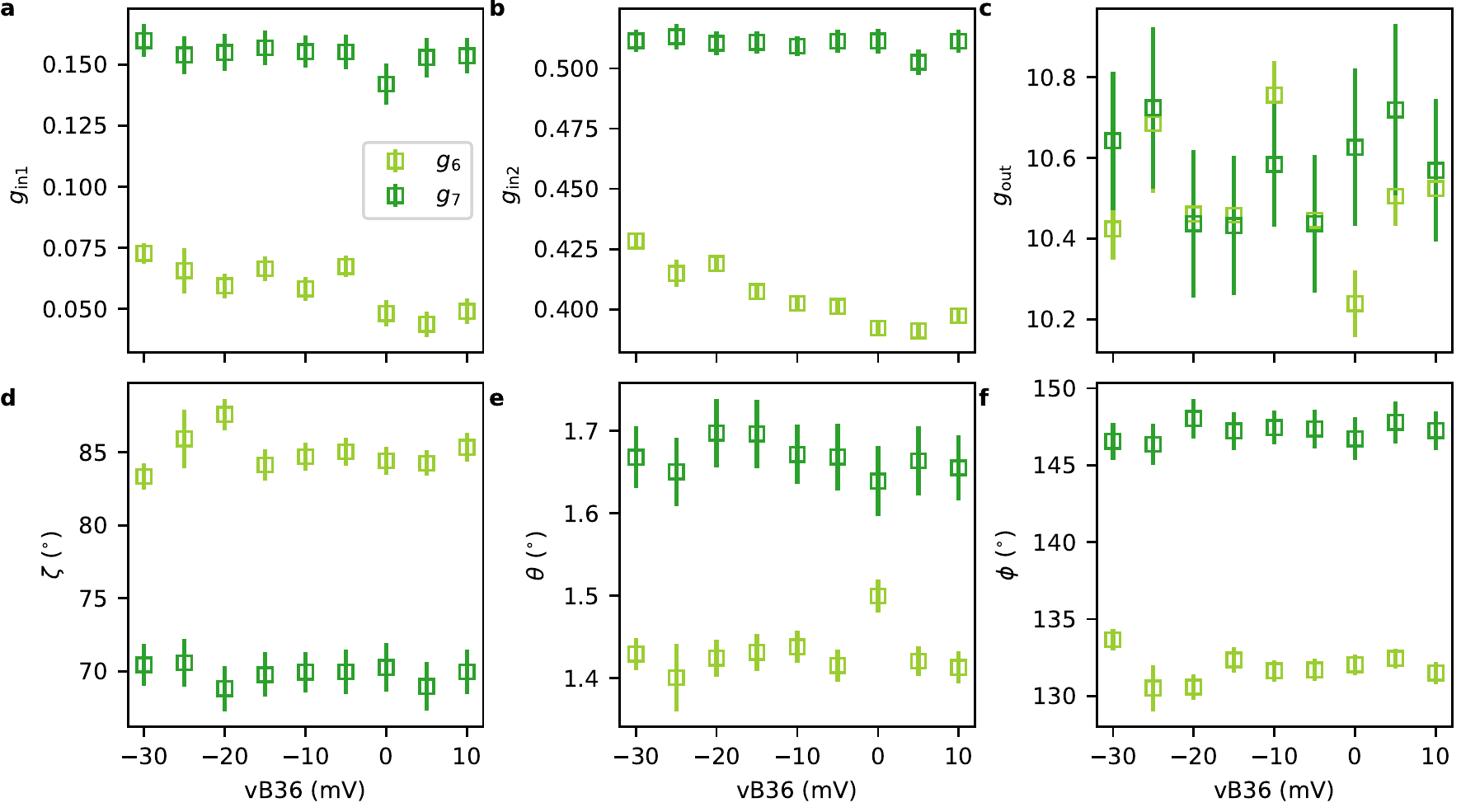}
    \caption{Gate tunability of $\tens{g}_6$ and $\tens{g}_7$ with interdot barrier B36 of device B. \figletter{a-f} Fitted principal axis \gtens{} values and $zyz$ Euler rotation angles for different barrier gate voltages vB36 applied to the barrier between QD3 and QD6.}
    \label{supfig:B36_tuna}
\end{supfigure*}

We note that a change of voltage for these barriers does not introduce a large rotation of the orientation angle $\zeta+\phi$.  In addition, only small changes of the tilt angle $\theta$ are observed.

\section{Spin-orbit vector fitting}
\label{sec:SO}

Following \cite{geyer_anisotropic_2024}, the single-particle Hamiltonian of a DQD with
spin-orbit interaction is given by:
\begin{align}
\label{eq:H1p}
\begin{split}
    \hat H_\mathrm{1P}
    &=
    \dfrac{\varepsilon}{2}\big(\ket{1}\bra{1}-\ket{2}\bra{2}\big)\\
    &{\quad\ } +\dfrac{\muB}{2} \vct{\hat\sigma} \cdot \left(\ket{1}\bra{1}\tens{g}_\mathrm{1}+\ket{2}\bra{2}\tens{g}_\mathrm{2}\right) \vct{B}\\
    &{\quad\ } +\of{\big[t_0 \ket{1}\bra{2} - i t_\mathrm{SO}\vct{n}_\mathrm{SO}\cdot\vct{\hat\sigma}\ket{1}\bra{2} \big] + h.c. }
\end{split}
\end{align}
where $\ket{i}$ is the ground orbital state of QD $i$, $\vct{\hat\sigma} = [\hat{\sigma}_x,\hat{\sigma}_y,\hat{\sigma}_z]$ is the spin vector operator with $\hat{\sigma}_z=\ket{\uparrow}\bra{\uparrow}-\ket{\downarrow}\bra{\downarrow}$, $t_0 = t_\mathrm{c}\cos\theta_\mathrm{SO}$ and $t_\mathrm{SO} = t_\mathrm{c}\sin\theta_\mathrm{SO}$) are the spin-conserving and spin-flip tunnel couplings, $\vct{n}_\mathrm{SO}$ the spin-orbit unit vector describing the direction of the generated spin-orbit field, $\varepsilon$ the detuning in energy units, and $h.c.$ denotes the Hermitian conjugate.
The two-particle Hamiltonian is constructed using the single-particle Hamiltonian from \eqnref{eq:H1p}:
\begin{align}
\label{eq:H2p}
    \begin{split}
        \hat{H}_\mathrm{2P}
        &=
        \hat{H}_\mathrm{1P} \otimes \hat{I} + \hat{I} \otimes \hat{H}_\mathrm{1P}\\
        &\quad \ +U_\mathrm{H}\big(\ket{1}\bra{1}\otimes\ket{1}\bra{1}+\ket{2}\bra{2}\otimes\ket{2}\bra{2}\big)
    \end{split}
\end{align}
with Hubbard charging energy $U_\mathrm{H}$. Since the particle-exchange-antisymmetric wavefunctions of \eqnref{eq:H2p} must obey Pauli's exclusion principle, we project the (symmetric) Hamiltonian to the following antisymmetric basis states
\begin{subequations}
\begin{align}
    \ket{\text{S}(2,0)} &=
    \frac{1}{\sqrt{2}} \of{\ket{1{\uparrow}}\otimes\ket{1{\downarrow}} - \ket{1{\downarrow}}\otimes\ket{1{\uparrow}}} ,\\
    \ket{ss'} &=
    \frac{1}{\sqrt{2}} \of{\ket{1s}\otimes\ket{2s'} - \ket{2s'}\otimes\ket{1s}}\quad \mathrm{for} \quad s,s'\in\{\uparrow,\downarrow\} .
\end{align}
\end{subequations}

The projection leads to the low energy two-particle Fock space Hamiltonian:
\begin{equation}
    \hat{H}_{5\times5}
    =
    \begin{bmatrix} 
    U_\mathrm{H}+\varepsilon & \tso\left(\nso{y}-i \nso{x}\right)&t_0 + i\tso \nso{z}&t_0 - i \tso\nso{z}&\tso\left(\nso{y}+i\nso{x}\right)\\
    \tso\left(\nso{y}+i\nso{x}\right)& \zee{z}{1}+\zee{z}{2}& \zee{x}{2}-i\zee{y}{2}& -\zee{x}{1}+i\zee{y}{1}&0\\
    t_0-i\tso\nso{z} & \zee{x}{2}+i\zee{y}{2}&\zee{z}{1}-\zee{z}{2}&0&\zee{x}{1}-i\zee{y}{1}\\
    t_0 + i\tso\nso{z}&-\zee{x}{1}-i\zee{y}{1}&0&-\zee{z}{1}+\zee{z}{2}&-\zee{x}{2}+i\zee{y}{2}\\
    \tso\left(\nso{y}-i\nso{x}\right)& 0 & \zee{x}{1}+i\zee{y}{1}&-\zee{x}{2}-i\zee{y}{2}&-\zee{z}{1}-\zee{z}{2}
    \end{bmatrix} ,
\end{equation}
where the order of the basis states is $\left\{ \ket{\text{S}(2,0)},\ket{\uparrow\uparrow},\ket{\uparrow\downarrow},\ket{\downarrow\uparrow},\ket{\downarrow\downarrow}\right\}$ and we defined the Zeeman vectors as $\vct{\varDelta}_i = \frac{1}{2} \muB \tens{g}_{\mathrm{Q}i} \vct{B}$.

The experiment is simulated by performing two subsequent time evolutions to describe the ramp from a $(2,0)$ to a $(1,1)$ charge state, as well as the return ramp. The return probability of a blocked state is given by:
\begin{equation}
\label{eq:prob}
    P_\mathrm{blocked}
    =
    1-\left|\bra{\text{S}(2,0)}\hat{U}(t_\text{ramp out})\hat{U}(t_\text{ramp in})\ket{\text{S}(2,0)}\right|^2
\end{equation}
where $\ket{S(2,0)}$ is used both as an initial state and as a measurement projector. The propagator in \eqnref{eq:prob} is 
\begin{equation}
    \hat{U}(t) = \mathcal{T}\Exp{-i\int^t_0 \D{t'} \hat{H}_{5\times5}(\varepsilon(t'))/\hbar} ,
\end{equation}
with $\varepsilon(t)$ is a piecewise linear ramp in time and where $\mathcal{T}$ denotes the time ordered integral.

The tunnel coupling and lever arm to convert the voltage detuning to an energy detuning are extracted from a separate spin funnel measurement. In addition, the measured \gtenss{} are input into the model. As only in-plane magnetic field directions are studied, we assume that the spin-orbit vector is restricted to the in-plane components. The spin-orbit vector can therefore be parametrized with a single angle $\gamma_\mathrm{n}$, where $\vct{n}_\mathrm{SO}=\left[\cos{\gamma_\mathrm{n}},\sin{\gamma_\mathrm{n}},0\right]$. Furthermore, the two tunnel couplings are not independent since $\tso=t_0\tan{\theta_\mathrm{SO}}$, where $\theta_\mathrm{SO} \propto l_\text{SO}^{-1}$ is inversely proportional to the spin-orbit length. 
The spin-orbit parameters ($\theta_\mathrm{SO}$, $\gamma_\mathrm{n}$) are determined by calculating map plots (\figrefs{supfig:ST12}{supfig:ST67}, panels c,d) corresponding to the experimental data (\figrefs{supfig:ST12}{supfig:ST67}, panel a) and calculating the least squares cost functions (\figrefs{supfig:ST12}{supfig:ST67}, panel b). The calculated cost functions have multiple minima with regard to the spin-orbit parameters and are very sensitive with regard to the input tunnel coupling. Therefore, we additionally add a qualitative comparison of all the minima (\figrefs{supfig:ST12}{supfig:ST67}, panels c,d) to ensure that the features of the measured data are captured by the model with these parameter sets. Moreover, we verify the spin-orbit parameters by also comparing  simulation and measurement for a ramp experiment with a variable ramp in and diabatic ramp out (\figrefs{supfig:ST12}{supfig:ST67}, panels e,f). 

\begin{supfigure*}
    \centering
    \includegraphics[width=\textwidth]{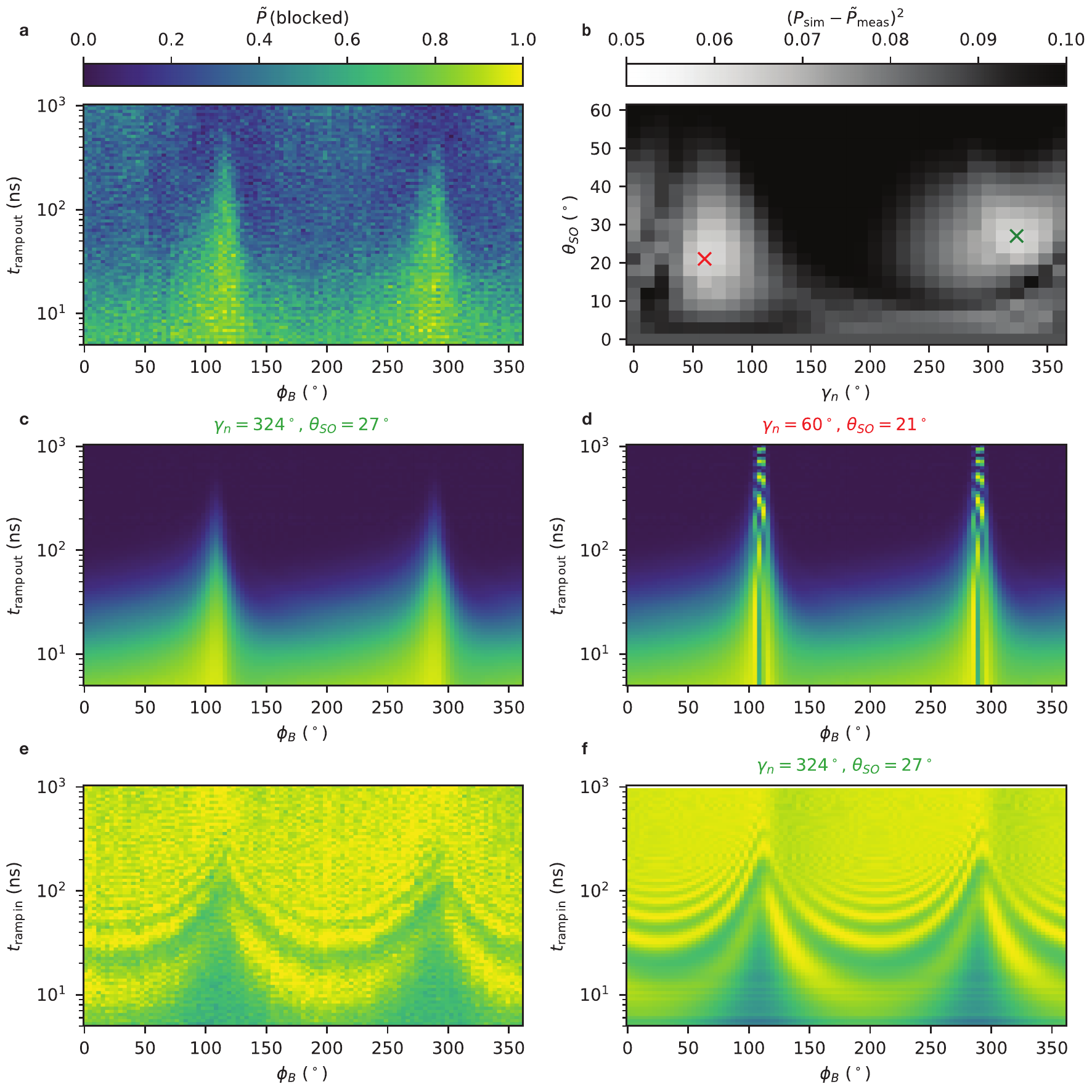}
        \caption{Spin-orbit field of DQD 12. \figletter{a}~Measured magnetic field angle dependence of the return probability of a blocked spin state for a ramp experiment with $\tin=1\us$ and variable $\tout$, with an external magnetic field of $7\mT$ applied in-plane. The probability is normalized (cf.\ \ref{sec:probability}). \figletter{b}~Comparison of measured and modeled return probability via cost function for varying spin-orbit strengths ($\theta_\mathrm{SO}$) and spin-orbit vectors $\vct{n}_\mathrm{SO}(\gamma_n)$ averaged for magnetic field strengths 5, 7, 10 and $20 \mT$. \figletter{c}~Simulation of the return probability at the absolute minimum (green marker in panel b) of the cost function. \figletter{d}~Simulation of the return probability at the local minimum (red marker in panel b) of the cost function. \figletter{e}~Measured magnetic field angle dependence of the return probability of a blocked spin state for a ramp experiment with variable $\tin$ and $\tout=1\ns$ with an external magnetic field of $7\mT$ applied in-plane. \figletter{f}~Simulation corresponding to the measurement in panel e with the same spin-orbit parameters as used for the simulation of panel c.}
    \label{supfig:ST12}
\end{supfigure*}
\begin{supfigure*}
    \centering
    \includegraphics[width=\textwidth]{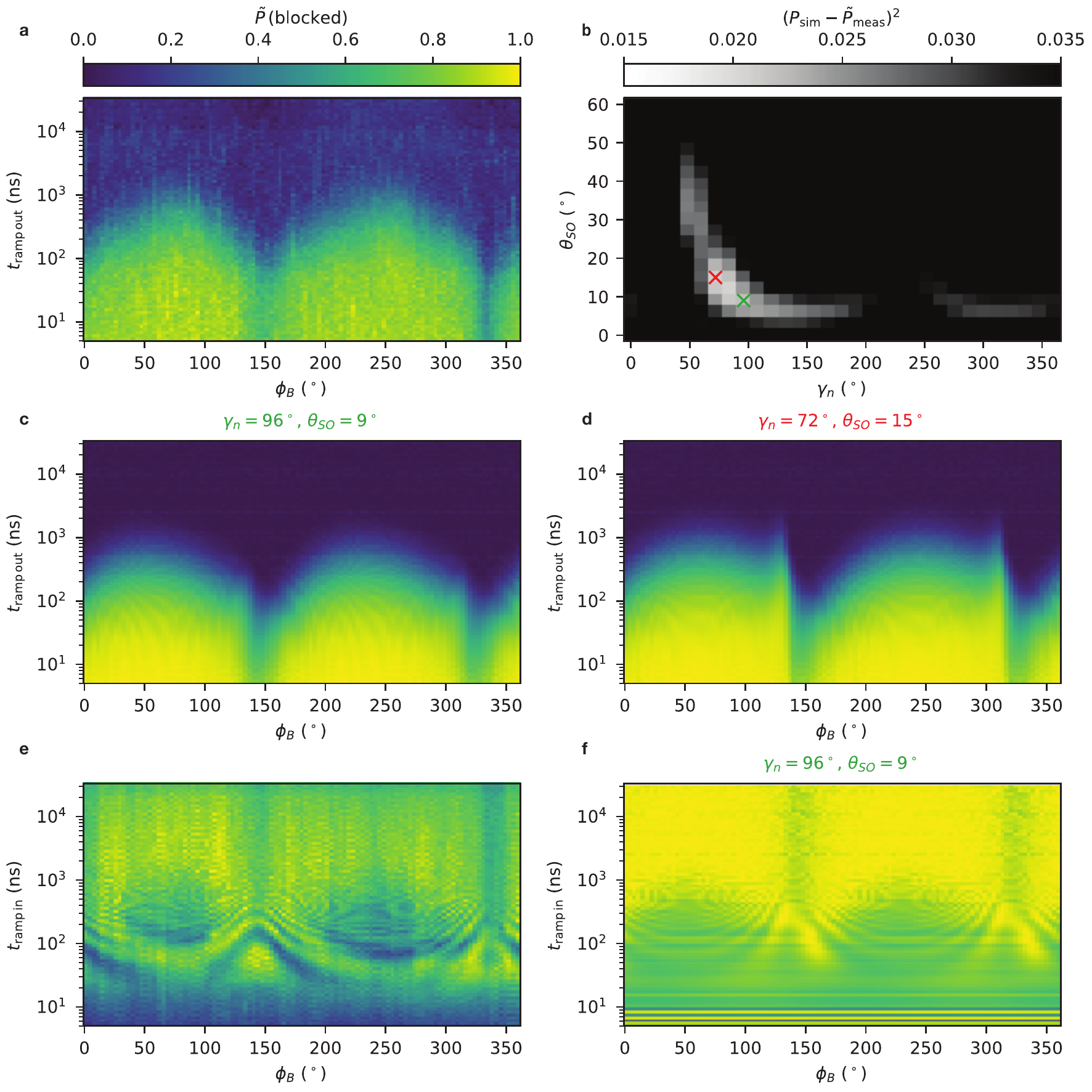}
        \caption{Spin-orbit field of DQD 45. \figletter{a}~Measured magnetic field angle dependence of the return probability of a blocked spin state for a ramp experiment with $\tin=30\us$ and variable $\tout$ with an external magnetic field of $7\mT$ applied in-plane. The probability is normalized (cf.\ \ref{sec:probability}). \figletter{b}~Comparison of measured and modeled return probability via cost function for varying spin-orbit strengths ($\theta_\mathrm{SO}$) and spin-orbit vectors $\vct{n}_\mathrm{SO}(\gamma_n)$ averaged for magnetic field strengths 5,7,10 and $20 \mT$. \figletter{c}~Simulation of the return probability at the with spin-orbit field parameters (green marker in panel b) near the minimum of the cost function. The parameters are chosen to get the best match to measured features in the probability map. \figletter{d}~Simulation of the return probability at the absolute minimum (red marker in panel b) of the cost function. \figletter{e}~Measured magnetic field angle dependence of the return probability of a blocked spin state for a ramp experiment with variable $\tin$ and $\tout=1\ns$ with an external magnetic field of $7\mT$ applied in-plane. \figletter{f}~Simulation corresponding to the measurement in panel e with the same spin-orbit parameters as used for the simulation of panel c.}
    \label{fig:ST45}
\end{supfigure*}
\begin{supfigure*}
    \centering
    \includegraphics[width=\textwidth]{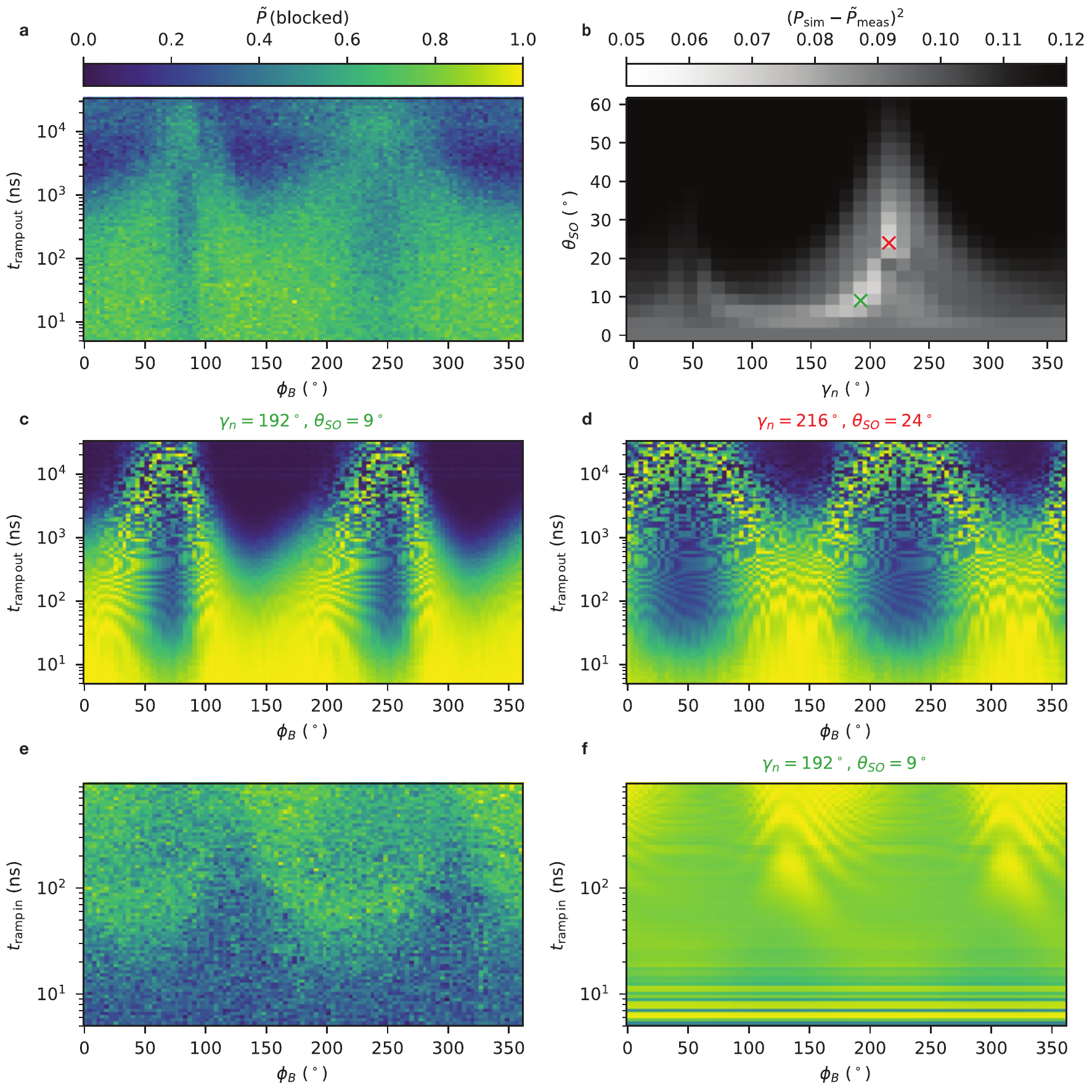}
        \caption{Spin-orbit field of DQD 67. \figletter{a}~Measured magnetic field angle dependence of the return probability of a blocked spin state for a ramp experiment with $\tin=30\us$ and variable $\tout$ with an external magnetic field of $7\mT$ applied in-plane. The probability is normalized (cf.\ \ref{sec:probability}). \figletter{b}~Comparison of measured and modeled return probability via cost function for varying spin-orbit strengths ($\theta_\mathrm{SO}$) and spin-orbit vectors $\vct{n}_\mathrm{SO}(\gamma_n)$ averaged for magnetic field strengths 5, 7, 10 and $20\mT$. \figletter{c}~Simulation of the return probability at the with spin-orbit field parameters of the local minimum (green marker in panel b) of the cost function. The parameters are chosen to get the best match to measured features in the probability map. \figletter{d}~Simulation of the return probability at the absolute minimum (red marker in panel b) of the cost function. \figletter{e}~Measured magnetic field angle dependence of the return probability of a blocked spin state for a ramp experiment with variable $\tin$ and $\tout=1\ns$ with an external magnetic field of $7\mT$ applied in-plane. \figletter{f}~Simulation corresponding to the measurement in panel e with the same spin-orbit parameters as used for the simulation of panel c.}
    \label{supfig:ST67}
\end{supfigure*}

\section{Probability normalization}
\label{sec:probability}
Comparing the data from the ramp experiments with the simulated return probability requires a mapping of current to probability. Due to large noise on the sensor current, it is not possible to distinguish the blocked and unblocked states in single shot measurements. To map to a probability, the minimal and maximal current signals are approximated to correspond to 0 and 1 respectively. This approximation assumes that the ideal adiabatic ramp in and ramp out instances can be achieved in the same scan and neglects any mixing due to the duration of the ramp. In addition, it assumes that for the shortest ramp times a purely diabatic ramp is achieved. 
As this is a very coarse approximation, a quantitative comparison between the measured and simulated data has only limited validity. To determine whether the cost-function based model evaluation is sensitive to the exact range of the measured probability, the spin-orbit parameters of the minimum of the cost function are calculated for varying probability ranges sweeping both the minimal and maximal measured probabilities (\figref{supfig:ST_probscale}). The obtained spin-orbit parameters mostly match the previously considered values. The few outliers can be eliminated by a qualitative comparison of the modeled probability with the measurement. 

\begin{supfigure*}
    \centering
    \includegraphics[width=\textwidth]{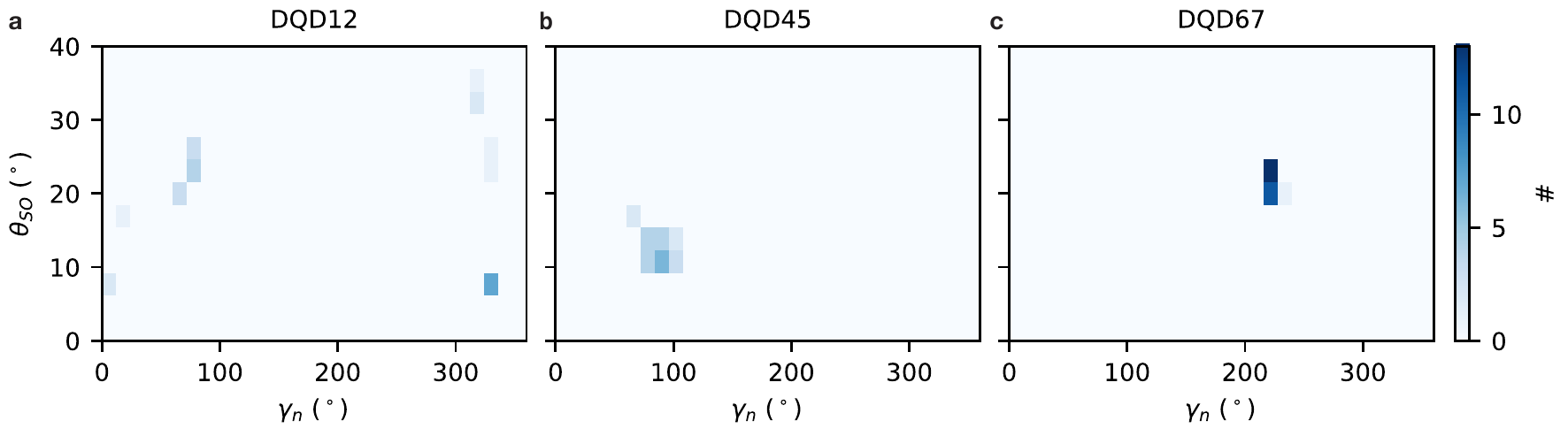}
        \caption{Spin-orbit parameters of cost function minima for different probability mapping of the measured current signal. Both the minimal and maximal measured probabilities are varied $\min(P)\in\left[0,0.2\right]$ and $\max(P)\in\left[0.8,1\right]$ for DQD12 \figletter{a}, DQD45 \figletter{b}~and DQD67 \figletter{c}. The marker color intensity scales with number of occurrences.}
    \label{supfig:ST_probscale}
\end{supfigure*}

\end{document}